\documentclass[twocolumn,showpacs,aps,superscriptaddress]{revtex4}

\usepackage{float} 
\usepackage{graphicx} 
\usepackage{amsmath}   
\usepackage{amssymb}   

\input psfig.sty

\def\bea{\begin{eqnarray}}
\def\eea{\end{eqnarray}}
\def\ben{\begin{equation}}
\def\een{\end{equation}}
\def\benu{\begin{enumerate}}
\def\enu{\end{enumerate}}

\def\n{n}

\def\sss{\scriptscriptstyle\rm}

\def\l{^\lambda}

\def\1var{(\bx_1...\bx\N)}

\def\bx{{x}}

\def\x{_{\sss X}}
\def\c{_{\sss C}}

\def\xc{_{\sss XC}}

\def\N{_{\sss N}}

\def\ee{_{\rm ee}}

\def\sph_int{ {\int d^3 r}}

\input rotate
\input epsf

\def\vec#1{{\mathbf #1}}

\begin{document}

\title{Describing static correlation in bond dissociation by Kohn-Sham \\ density functional theory}

\author{M. Fuchs}
\affiliation{
Fritz-Haber-Institut der Max-Planck-Gesellschaft,
Faradayweg 4-6, D-14195 Berlin, Germany}
\affiliation{Unit{\'e} PCPM, Universit{\'e} Catholique de Louvain,
1348 Louvain-la-Neuve, Belgium}
\author{Y.-M. Niquet}
\affiliation{Unit{\'e} PCPM, Universit{\'e} Catholique de Louvain,
1348 Louvain-la-Neuve, Belgium}
\affiliation{D{\'e}partement de Recherche Fondamentale sur la Mati{\`e}re Condens{\'e}e, SP2M/L\_Sim, CEA Grenoble, 38054 Grenoble Cedex 9, France}
\author{X. Gonze}
\affiliation{Unit{\'e} PCPM, Universit{\'e} Catholique de Louvain,
1348 Louvain-la-Neuve, Belgium}
\author{K. Burke}
\affiliation{Dept. of Chemistry and Chemical Biology, Rutgers
University, 610 Taylor Road, Piscataway, NJ 08854} 
\date{\today}

\begin{abstract} 
We show that density functional theory within the RPA (random 
phase approximation for the exchange-correlation energy)
provides a correct description of bond dissociation in H$_2$
in a spin-restricted Kohn-Sham formalism, i.e. without artificial 
symmetry breaking. We present accurate adiabatic 
connection curves both at equilibrium and beyond the Coulson-Fisher 
point. The strong curvature at large bond length implies important
static (left-right) correlation, justifying modern hybrid 
functional constructions but also demonstrating their limitations.
Although exact at infinite and accurate around the equilibrium
bond length, the RPA dissociation curve displays unphysical
repulsion at larger but finite bond lengths. Going beyond the
RPA by including the exact exchange kernel (RPA+X), we find
a similar repulsion. We argue that this deficiency is due to the 
absence of double excitations in adiabatic linear response theory.    
Further analyzing the H$_2$ dissociation limit we show that
the RPA+X is not size-consistent, in contrast to the RPA.
\end{abstract}

\date{\today}
\maketitle

\section{Introduction}	
Density functional theory~\cite{HK64,KS65,kh01a} (DFT) has proved 
to be a powerful method for calculating (and analyzing) 
the ground-state properties of molecular and condensed matter.
In its standard Kohn-Sham (KS) form, the density $n(\vec r)$
and total energy are constructed from the self-consistent
solution of one-electron equations where, in practice,
the exchange-correlation (XC) energy functional $E\xc[n]$ 
must be approximated.
Already the simple local-density approximation (LDA) and, more so, 
generalized gradient approximations (GGAs) can yield a remarkably 
realistic description of chemical bonds in solid and molecular 
systems. The state of the art is presently set by hybrid
functionals~\cite{bec93a,per96a} that admix a fraction of the exact (Fock) 
exchange energy with GGA exchange. Achieving on the average 
nearly chemical accuracy for bond dissociation energies, they
rival much more demanding post-Hartree-Fock configuration 
interaction or coupled cluster methods.

However there remain well-known conceptual limitations, with a 
clear practical significance, as exemplified by the following
paradigm situations.
$(i)$ Long-ranged Coulomb correlations between non overlapping
systems are not included in local functionals such as the LDA or 
GGAs (and thus also hybrids) but require fully nonlocal XC energy
functionals~\cite{dob94a,dob96a,koh98a}. Indeed LDA and GGAs perform 
at best erratically for van der Waals bonded systems \cite{eng00b,ggg02}.
$(ii)$ Scaling to the high-density or weakly-interacting regime, 
hybrid functionals do not properly recover the exact KS exchange
energy~\cite{ern97a}. This failure prominently concerns 
odd electron bonds, as exemplified by the H$^+_2$ molecule, where 
$100$\% exact exchange mixing and zero correlation energy would 
be needed~\cite{bal97a,sod99a}. 
$(iii)$ In systems with significant static (non-dynamical) 
correlation, LDA, GGA, and thus hybrid functionals underestimate 
the magnitude of the correlation energy~\cite{ern97b,gri97a}. 
This becomes particularly problematic for the dissociation of 
electron pair bonds where near degeneracy effects arise in the 
molecular wavefunction. A famous example is the dissociating 
H$_2$ molecule: the proper (singlet) KS ground state at larger 
bond lengths has much too high total energy for such functionals. 
Usually one works around the problem: performing a spin-unrestricted 
calculation, a second solution with reasonable, lower 
energy is obtained. This succeeds because exchange functionals can mimic
the static (i.e. long-ranged left-right) correlation and thus
compensate for an error of the LDA and GGA type correlation
functionals that describe effectively only dynamical correlation
coming from electron repulsion at short range~\cite{gri97a}.
Yet the spin-unrestricted KS molecular wavefunction artifically
breaks the symmetry of the dissociating molecule, displaying 
unphysical non-zero spin-polarization~\cite{per95a}. In fact 
{\em spurious} symmetry breaking has remained a much discussed 
drawback of spin-unrestricted DFT (and Hartree-Fock) calculations 
of molecular properties (see e.g. Refs. ~\cite{lee93a,gor99a,fil00a,pol02a}). 
 
Progress in each of the above noted respects may be achieved through 
orbital-dependent XC functionals expressed in terms of the 
(occupied and unoccupied) KS eigenstates, such as the 
Random Phase Approximation (RPA).
The RPA functional is the simplest realization of the adiabatic-connection 
fluctuation-dissipation formalism~\cite{lan75a} that defines 
a broad class of fully nonlocal XC functionals. 
In terms of the ``Jacob's ladder'' of density functional
approximations~\cite{perdew01p}, RPA is on the top
rung, putting it among the most general (but also the
most demanding) of present-day approximations.
It includes 
the exact KS exchange energy~\cite{kur99a} as well as long-ranged 
Coulomb correlations giving rise to dispersion forces~\cite{dob94a,dob00a}. 
This makes the RPA a natural starting point to address the above 
limitations $(i)-(iii)$ of existing functionals in a seamless 
and consistent way. In this paper we 
show that RPA functional is able to describe the strong static correlation 
in the dissociation of the H$_2$ bond in a {\em spin-restricted} 
KS formalism, without artificial symmetry breaking. The
transition from mostly dynamical to strong static
correlation is discussed in terms of the adiabatic connection. 
Analyzing the dissociation energy curve of H$_2$, we 
find that the RPA is accurate around the equilibrium bond 
length and yields, asymptotically, the correct dissociation 
into two H atoms. At intermediate distances the dissociation energy 
still displays an erroneous repulsion. Of course the RPA is just the first step 
in an ongoing systematic quest, with encouraging results for
small molecules~\cite{fur01a,fuc02a} as well as van der Waals 
bonded structures~\cite{dob00a,ryd03a,dio04a}. 
To achieve satisfactory accuracy globally it requires extensions, 
some of which are being examined already~\cite{koh98a,kur99a,yan00a,dob00b,dob02a,fuc02a}.

Last we would like to mention that the XC energy can be approximated also as 
functional of the one-electron density matrix, by making an appropriate ansatz 
for the exchange and correlation hole functions. For a particular 
such functional involving both the occupied and the unoccupied KS 
states, Gr\"uning {\it et al.}~\cite{gru03a} recently reproduced 
the entire dissociation energy curve of H$_2$, including proper 
dissociation. The validity of this approach is however still under 
debate~\cite{her03a}.

Our paper is organized as follows. In Sec.~\ref{sec:dilemma} we first 
recall the classic problem of stretched H$_2$ and its implications
in the context of different density functionals, and briefly 
discuss ways of dealing with it. We also summarize
the RPA equations. In Sec.~\ref{sec:dissociation} 
we examine the XC energy of H$_2$ in the framework of the
adiabatic connection, then,   
in Sec.~\ref{sec:results} we present and discuss 
our results for the RPA dissociation energy curve for H$_2$, and
comment on current limitations and ways to overcome them. 
The Sec.~\ref{sec:extensions} is devoted to
results obtained beyond the RPA. Section
~\ref{sec:summary} summarizes our conclusions.

\section{The problem of stretched H$_2$ in DFT}
\label{sec:dilemma}

A long-standing problem confronting all single-determinant
calculations is that of stretched H$_2$, representative
for the dissociation of electron pair bonds in general~\cite{kh01a,per95a,bae01a}. 
For {\em any} bond length $R$ the true (interacting) ground state 
is a singlet~\cite{kol60a}, with equal spin-up and spin-down
densities, and the true (noninteracting) KS ground-state corresponds to a
single Slater determinant $\Psi^{\rm KS}=|\sigma_g\overline\sigma_g|$
made up from the bonding $\sigma_g$ molecular orbital. 
The spin-restricted LDA, GGA, and hybrid functionals (as
well as Hartree-Fock) correctly yield such a ground state,
with reasonable total energy, around the equilibrium 
bond length $R_0$. As a matter of fact, the interacting
ground state is also mainly of $|\sigma_g\overline\sigma_g|$
nature around $R\approx R_0$. However, as the bond length $R$
is increased toward $R\rightarrow\infty$, $\Psi^{\rm KS}$
no longer resembles the interacting ground state wavefunction
of the molecule.
Asymptotically the latter assumes the familiar Heitler-London
form and puts precisely one electron on each of the two
hydrogen atoms, completely suppressing number fluctuations
and describing two free hydrogen atoms, i.e. 
$\mbox{H}^{\cdot}{\cdots}\,\mbox{H}^{\cdot}$~\cite{per95a}. 
By contrast $\Psi^{\rm KS}$ is half contaminated by 
ionic contributions of the form $|s_a\overline s_a|$ and $|s_b\overline s_b|$ 
($s_a$ and $s_b$ stand for the $1s$ orbitals centered
on hydrogen A and B respectively), in effect describing 
the stretched H$_2$ as 
$\frac{1}{2}(\mbox{H}^{\cdot}{\cdots}\,\mbox{H}^{\cdot}) + \frac{1}{2}(\mbox{H}^+{\cdots}\,\mbox{H}^{-})$
~\cite{com:ymn}.
Optimizing e.g. within spin-restricted GGA, the $\sigma_g$ orbital 
does not become a linear combination of the $1s$ hydrogenic orbitals
but gets much too diffuse, in order to avoid the H$^{-}$ contribution. 
Hence the dissociation energy of stretched H$_2$ is severely 
overestimated, its total energy lying much above the one of 
two free hydrogen atoms (as illustrated later by our Fig.~\ref{fig:pes}).
On the other hand, a second solution with {\em lower} (and
reasonable) energy may be obtained for bond lengths beyond 
the so-called Coulson-Fisher point~\cite{gun76a,cou} by breaking the 
spin-symmetry and localizing on one hydrogen atom the ``spin-up'' 
electron and on the other the ``spin-down'' electron. This is what 
is obtained, in practice, by performing spin-unrestricted L(S)DA, 
GGA, or hybrid functional (as well as unrestricted Hartree-Fock) 
calculations. 

This situation for approximate XC functionals embodies
the well-known spin symmetry dilemma for dissociating H$_2$:
restricted KS schemes yield the proper symmetry adapted
ground-state but poor total energies, while unrestricted
KS schemes yield very reasonable total energies but 
qualitatively wrong spin densities~\cite{per95a}.

\subsection{Local, semilocal, and hybrid functionals}
\label{sec:modernDFT}

A discussion of this dilemma might appear academic, since
in most cases one simply calculates the chemically bonded
system, and subtracts from it the energies of the isolated
atoms, without ever watching the molecule dissociate.
However, at the origin of the problem is the inability of
present XC functionals to properly capture long ranged
left-right correlation that eventually appears when a 
molecule dissociates~\cite{leeuwen96}.
Such static (or non-dynamical) correlation is present in many 
real molecules even in or near the ground-state geometry. As 
a matter of fact LDA and GGA calculations only mimic it 
through their exchange component while their correlation 
component accounts for (short-ranged) dynamical correlation 
only (see e.g. Ref.~\cite{gri97a}). In fact their performance 
depends crucially on the well-known cancellation of errors 
between exchange and correlation~\cite{ern97a}. 
This compensation is of course neither perfect nor universal, 
as indicated by the on average overestimated atomization 
energies, particularly for multiply-bonded species like N$_2$, 
but also by often underestimated reaction energy barriers 
(see e.g. Ref.~\cite{fil02a}).

Hybrid functionals~\cite{bec93a,per96a} can provide a useful 
remedy and are usually more accurate than LDA or GGAs alone.
Yet they still rely on a similar philosophy: the admixing of a 
certain {\em fixed} fraction of the (possibly spin-unrestricted) 
exact exchange energy can be understood to improve the description 
of static correlation while dynamical correlation is still 
described by (semi-) local LDA or GGA functionals. The errors in 
the exchange and correlation energies do not sufficiently cancel 
in cases where either exchange (paradigm: H$_2^+$) or static 
correlation (paradigm: dissociated H$_2$) prevail, so that 
eventually hybrid functionals can become inadequate too~\cite{ern97b}. 
For dissociating electron pair bonds, sufficiently negative 
exchange energies are obtained only by resorting 
to spin-unrestricted exchange, i.e. by an unphysical breaking of 
the symmetry of the KS molecular wavefunction.
At which bond length the spin-unrestricted solution becomes
energetically preferable depends on the functional. 
Hybrid functionals are more prone to break symmetry with an increased
admixing of exact exchange (see Ref.~\cite{bau96a}).
Artificially symmetry broken solutions from spin-unrestricted
methods may also yield unphysical molecular properties, despite 
providing a lower energy solution and appearing better 
variationally~\cite{gor99a}. We return to the concept of hybrid 
functionals in Sec.~\ref{sec:ac}.

\subsection{Functionals of occupied and unoccupied KS states}
\label{sec:compatibleXC}

Although the physical origin of the above difficulties is 
clear it is far from simple to correct for them, while 
retaining low computational cost. A variety of approaches have 
been applied, mostly along the same lines as a traditional 
restricted Hartree-Fock calculation would be corrected, such 
as spin-unrestricted~\cite{gun76a}, multi-reference~\cite{gra98a,gri99a}, 
or ensemble-referenced Kohn-Sham schemes~\cite{bae99f,fil00a}. 
Although often useful, these suffer from similar difficulties as 
in Hartree-Fock, namely that different approaches work for 
different situations. 

A more satisfying approach is to develop a more demanding
but non-empirical scheme. Functionals involving occupied
and unoccupied (virtual) KS orbitals offer this possibility
as recently shown by Baerends and coworkers~\cite{bae01a,gru03a}. 
A well-defined starting point is the exact exchange formalism 
(EXX) in Kohn-Sham DFT~\cite{goe94a} which then must be 
complemented with a compatible correlation functional, i.e., 
one that properly respects the weakly-correlated, 
exchange-only limit (e.g. producing zero correlation 
energy in one-electron systems such as H$_2^+$) and accurately 
interpolates to the strongly-coupled regime to be discussed 
in Sec.~\ref{sec:ac}. The RPA XC functional, discussed next 
in Sec.~\ref{sec:rpa} represents a possible first step into 
this direction.

\subsection{Functionals from linear response: RPA and beyond}
\label{sec:rpa}

The combination of the random-phase approximation (RPA) with
DFT was discussed as early as the late 70's~\cite{lan75a}, then
for the uniform electron gas. The idea is to start from the 
non-interacting KS response function and dress it with a
(scaled) Coulomb interaction $\lambda{\hat v}\ee$,
and eventually also with an exchange-correlation 
kernel of time-dependent DFT (TDDFT)~\cite{pet96a}. This 
yields the response function of an interacting system, which 
by way of the fluctuation-dissipation theorem gives the corresponding
pair-correlation function and thus the electron-electron interaction energy.
The XC energy is last obtained through an integration of the 
electron-electron interaction energy along an adiabatic path connecting the non-interacting KS system ($\lambda=0$) to the interacting one ($\lambda=1$, see Sec.~\ref{sec:ac}).

Working in the imaginary frequency ($iu$) domain, the basic 
equations (in Hartree atomic units) are as follows. 
Starting from a KS ground-state with eigenstates 
$\{\phi_{i\sigma}[n],\epsilon_{i\sigma}[n]\}$, 
functionals of the density $n$, one constructs the KS 
response
\begin{eqnarray}
 \chi^0(\mathbf r,\mathbf r';iu)
         & = & \displaystyle{
               \sum_{\sigma,k,l}
               \frac{(\gamma_{k\sigma}\!-\!\gamma_{l\sigma})}
                    {iu - (\varepsilon_{l\sigma} - \varepsilon_{k\sigma})}
                    }
\nonumber
\\
        &&     \times \phi_{k\sigma}^{\ast}(\mathbf r) \phi_{l\sigma}(\mathbf r)
                      \phi_{l\sigma}^{\ast}(\mathbf r')\phi_{k\sigma}(\mathbf r')
\quad ,
\label{eq:chiks}
\end{eqnarray}
with $\gamma_{i\sigma}=1$ for occupied and $\gamma_{i\sigma}=0$ for unoccupied 
KS states. The interacting response function $\chi^{\lambda}$ at coupling strength 
$\lambda$ follows from the Dyson-type screening equation
\begin{equation}
\label{eq:dyson}
\chi^{\lambda}(iu) = \chi^0(iu) [1 - \left(\lambda v_{\rm ee} + f\xc^{\lambda}(iu)\right)\chi^0(iu)]^{-1}
\quad,
\end{equation}
where $v_{\rm ee}=1/|\vec r-\vec r'|$ is the Coulomb repulsion and $f\xc^{\lambda}(iu)$ 
stands for the XC kernel of TDDFT (matrix notation $A =: A(\vec r,\vec r')$ and 
$AB=: \int d^3r'' A(\vec r,\vec r'')B(\vec r'',\vec r')$ is implied here). 
The fluctuation-dissipation theorem and the coupling strength integration (see Sec.~\ref{sec:ac}) 
finally yield an exact expression for the XC energy:
\begin{eqnarray}
E\xc[n] &=& \!
       -\frac{1}{2} \int_0^{1}d\lambda \int d^3r d^3r'
        \frac{1}{|\mathbf r - \mathbf r'|}
        \quad\quad\quad\quad\quad\quad
\nonumber
\\
  && \times\! \left[\left\{\int_0^{\infty}\!\frac{du}{\pi}\,
                                \chi^{\lambda}(\mathbf r,\mathbf r';iu)
                                \!
			\right\}
                        + n(\mathbf r)\delta(\mathbf r\! -\! \mathbf r')
                        \right]
                        ,
\label{eq:acfd}
\end{eqnarray}
called the adiabatic-connection fluctuation-dissipation theorem (ACFDT) 
for the XC energy. Through approximations for $f\xc^{\lambda}$, fully nonlocal 
ACFDT XC functionals can be generated in practice. In general these will be
orbital-dependent functionals involving, through $\chi^0$ (and eventually $f\xc^{\lambda}$), 
both occupied and unoccupied KS states. The RPA is obtained by setting
$f\xc^{\lambda}=0$, i.e. neglecting all XC contributions to screening
in Eq.~\ref{eq:dyson}. Approximating $f\xc^{\lambda}$ by the (exact) 
exchange kernel of TDDFT~\cite{pet96a} defines the RPA+X functional.
In the ACFDT formula (Eq.~\ref{eq:acfd}), because of the integral over
the coupling strength, an approximation to $\chi^{\lambda}$ of a given order in
$\lambda$, produces an approximation to $E\xc[n]$ to the next highest order.
Thus, to zeroth order, $\chi^0$ inserted in Eq.~\ref{eq:acfd} yields
the exact exchange energy, $E\x[n]$, the first-order energy. The RPA yields an
approximation to $E\xc$ that contains all orders of $\lambda$, but misses
contributions from second-order onward. 
In RPA+X on the other hand $f\xc^{\lambda}$ (and thus $\chi^{\lambda}$)
are treated exactly up to first-order. In turn the RPA+X yields an
approximation to $E\xc[n]$ that is exact up to second-order but, in
an approximate way, again includes also all high orders of $\lambda$.
Being exact to second-order, the RPA+X produces the exact initial slope of the 
adiabatic connection (to be discussed in detail in Sec.~\ref{sec:ac})
as given by second-order G\"orling-Levy perturbation theory (GL2)~\cite{goe93a}.

While the RPA is expected to be quite accurate for (iso-electronic) total energy 
differences, it tends to underestimate absolute correlation energies~\cite{kur99a}.
Indeed, the short-range behavior of $\chi^{\textrm{RPA}}$ is incorrect, because
the electrons only respond to the averaged density fluctuations. In particular,
the (spurious) ``self-response'' of an electron to its own contribution to these 
density fluctuations is the most important source of error in few electron systems. 
It is responsible for a (spurious) non-zero self-correlation energy in one-electron 
systems like the H atom.
This deficiency of the RPA may well be corrected by a local-density functional 
$E^{\rm sr-LDA}\c[n]$ designed for the purpose~\cite{kur99a}, defining the so-called ``RPA+'' functional 
$E\xc^{\rm RPA+}[n]=E\xc^{\rm RPA}[n]+E\c^{\rm sr-LDA}[n]$\,.
As expected in Refs.~\cite{kur99a,yan00a} and confirmed in Refs.~\cite{fur01a,fuc02a},
local or semilocal short-ranged corrections have rather minor effects on molecular 
dissociation energies, yet nicely correct for spurious RPA self-response problem in 
the H and He atoms~\cite{fuc02a}. For these reasons we focus in this paper on the 
{\em difference} of the total energies and XC energies between the molecule and
the isolated atoms.

Compared to likewise orbital-dependent Meta-GGAs~\cite{bec98a,scu98a,tao03a} 
and hybrid functionals, the RPA involves both {\em occupied} and {\em unoccupied} KS 
states. Its computational cost is quite severe (a factor 
$10^2\ldots 10^3$ compared to a GGA approach), but its promise 
is to tackle the nonlocality of exchange and correlation on an 
equal footing, leading to a seamless description from the 
chemically-bonded to the dissociated regime, including van
der Waals interactions not accounted for in GGAs or hybrids.

Recent calculations for small molecules by Furche~\cite{fur01a}
found that the RPA describes the chemical bonds with similar but
not better accuracy as modern GGAs. Although this appears to be 
disappointing, we note that the RPA is in fact accurate for H$_2$ 
and the difficult case of Be$_2$~\cite{fuc02a} where LDA and GGA 
fail. Also there has been significant progress in building 
XC functionals on top of the RPA that include the dispersion 
forces between layered solid-state structures such as jellium 
slab models~\cite{dob00a} and graphite~\cite{ryd03a}, and between 
atoms or molecules~\cite{dio04a}.
We further point out that the RPA is just a ready realization of 
a much broader class of functionals based on the adiabatic-connection 
fluctuation-dissipation theorem which offers various options
for improvements~\cite{fuc02a,yan00a,dob00b,dob02a}. Moreover the 
RPA is amenable to extensions derived in many-body Green 
function theory which itself is being very actively explored for 
total energy calculations~\cite{ggg02,alm99a,dah04a}, including 
H$_2$~\cite{hol90,ary01a,fuc03a}.

In the present study we do not perform self-consistent RPA calculations, 
these are computationally too costly at present (although formally perfectly
feasible, see Refs.~\cite{niq03a,niq03b}). Instead we evaluate
Eqs.~\ref{eq:chiks} - ~\ref{eq:acfd} {\it a posteriori} with the KS 
eigenstates taken from EXX calculations as explained further in 
Sec.~\ref{sec:dissociate}. Thanks to the variational principle for DFT 
total energies we expect the resulting RPA total energies to be tight 
upper bounds to the self-consistent RPA results. Indeed we found that 
our results for the RPA total energies remained virtually unchanged 
when we used the LDA~\cite{xclda} or PBE GGA~\cite{xcpbe} instead 
of the EXX for calculating the initial KS states.

\section{Analysis of the H$_2$ dissociation}
\label{sec:dissociation}

In this section and in the next one, we show that the RPA offers a
promising handle on the problem of dissociating electron
pair bonds as exemplified by H$_2$. In particular, we give 
the correct prescription for applying the scheme during 
dissociation and show that its remaining errors may well 
be correctable. Using it, we also provide the first accurate 
calculations of adiabatic connection curves as a system passes 
through its Coulson-Fisher point where the KS ground-state
for standard XC functionals bifurcates in a spin-symmetry
adapted and a lower energy but symmetry-broken solution.

\subsection{Adiabatic connection} 
\label{sec:ac}

To understand in detail how DFT handles static correlation,
one invokes the adiabatic connection~\cite{lan75a,gun76a},
already briefly mentioned in Sec.~\ref{sec:rpa}. One imagines
altering the strength of the electron-electron repulsion
by multiplying it by a constant $\lambda$, which varies between
0 and 1.  At the same time, the one-body potential is altered,
i.e., made a function of $\lambda$, so as to keep the
electron density fixed.  
This is a way of continuously connecting the noninteracting
KS system ($\lambda=0$) to the interacting physical system
($\lambda=1$).  More importantly, by
virtue of the Hellmann-Feynman theorem, one can write the
XC energy as an integral over purely potential energy:
\ben
E\xc[\n] = \int_0^1 d\lambda\  U\xc[\n](\lambda)
\label{AC}
\een
where $U\xc[\n](\lambda) = \langle \Psi\l[n] | {\hat v}\ee | \Psi\l[n] \rangle - U[\n]$.
Here $\Psi\l[\n]$ is the ground-state wavefunction at coupling strength
$\lambda$, ${\hat v}\ee$ is the Coulomb repulsion, and $U[\n]$ is the
Hartree energy. The integrand $U\xc[\n](\lambda)$ represents the
potential energy contribution to XC and makes up the adiabatic connection
curve. At the $\lambda=0$ end it 
corresponds to the exact Kohn-Sham exchange energy $E\x$~\cite{kur99a}, 
\begin{equation}
\label{eq:acex}
U\xc[\n](0)=E\x[\n]\quad,
\end{equation}
and has a (negative) initial slope, given by the correlation energy of second-order 
G{\"o}rling-Levy perturbation theory (GL2)~\cite{goe93a}
\begin{equation}
\label{eq:acgl2}
U\xc^\prime[\n](\lambda)=\left.\frac{d}{d\lambda} U\xc[\n](\lambda)\right|_{\lambda=0}= 
2\,E\c^{\rm GL2}[\n] 
	\quad.
\end{equation}
All $\lambda$-dependence rests in the correlation contribution
\begin{equation}
\label{eq:ucl}
U\c[\n](\lambda)=U\xc[\n](\lambda)-E\x[\n]\quad.
\end{equation}
At $\lambda=1$, $U\xc[n](\lambda)$ describes the XC potential energy
of the physical system, $U\xc[n]=:U\xc[\n](1)$, and, similarly,
the correlation potential energy, $U\c[n]=:U\c[\n](1)$.
Since correlation reduces the
electron-electron repulsion, the different energies are always ordered
as
\begin{equation}
E\x[n] \geq E\xc[n] \geq U\xc[n]
\quad.
\end{equation}
The $\lambda$-dependence for ACFDT type functionals (see Eq.~\ref{eq:acfd}), 
appearing through the response function $\chi^{\lambda}$,
is given by
\begin{equation}
\label{eq:acacfd}
U\c[\n](\lambda) =  
	-\frac{1}{2}\int_0^{\infty}\!\frac{du}{\pi}\,\,
	\mathrm{Tr}\left[ {v}\ee \left\{\chi^{\lambda}(iu)-\chi^{0}(iu)\right\}\right]
\quad,
\end{equation}
where $\mathrm{Tr}[A] =: \int d^3r A(\vec r,\vec r)$\,. The
full $\lambda$-curve may then be calculated thanks to Eq.~\ref{eq:ucl}.
For the LDA or GGA type functionals the $\lambda$-dependence
can be readily calculated from the exact relation~\cite{lev85c}
\begin{equation}
U\xc[\n](\lambda) = \frac{d}{d\lambda}\left(\lambda^2 E\xc[n_{1/\lambda}]\right)\quad,
\end{equation}
where the XC energy functional is evaluated at the scaled density 
$n_{1/\lambda}(\vec r)=:n(\vec r/\lambda)/\lambda^3$.
Hence analysis by adiabatic decomposition allows investigation into how 
well the different functionals perform, and why~\cite{ern96a,ern97a}.

Below we consider molecular dissociation energies, i.e. the difference
between the molecular and atomic total energies,
\begin{equation} 
\Delta E = E_{\textrm{tot}}[\mbox{molecule}] - E_{\textrm{tot}}[\mbox{atoms}]\quad,
\end{equation}
and analyze the exchange-correlation contributions by means 
of the analogously defined differences $\Delta E\xc$, $\Delta E\x$, 
and $\Delta U\xc(\lambda)$.

A useful measure of the correlation strength is given in Ref.~\cite{BEP97}.
Define the parameter $b$ by
\begin{equation}
E\xc[n] = b E\x[n] + (1-b) U\xc[n]
\quad.
\end{equation}
A simple interpretation of $b$ is given by the following geometrical
construction:
if the adiabatic curve were a horizontal line of value $E\x[n]$ running from $0$ to
$b$, and then dropped discontinuously to
another horizontal line of value $U\xc[n]$ running from $b$ to $1$, then
$b$ is that value of $\lambda$ that yields the correct $E\xc[n]$.
Thus, in the high density limit, where the adiabatic connection curve
is a straight line, $b$ is exactly $1/2$. On the other hand, for
strong static correlation, in which the adiabatic connection curve
drops rapidly to its final value, $b$ is close to zero.  One can also 
show~\cite{BEP97}
\begin{equation}
\label{eq:defb}
b = \frac{T\c[n]}{|U\c[n]|}\quad,
\end{equation}
where $T\c$ is the kinetic portion of the correlation energy
\begin{equation}
\label{eq:tc}
T\c[n]=E\c[n]-U\c[n]\quad.
\end{equation}
Thus small $b$ indicates that the correlation is indeed static, i.e.,
has a smaller fraction of kinetic to potential energy.

\begin{figure}
\includegraphics[clip=,width=0.9\linewidth]{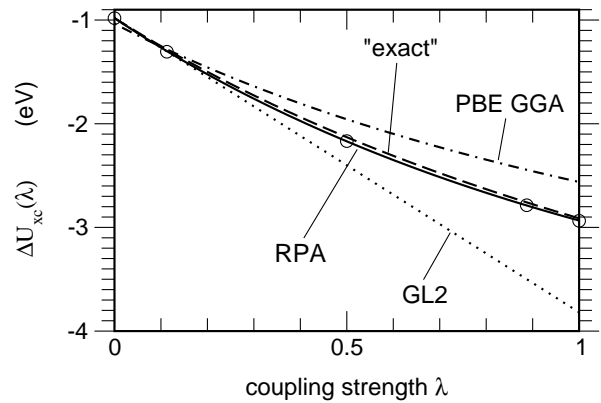} 
\caption{Adiabatic connection for H$_2$ at bond length $R=1.4$~bohr
within the RPA (solid line) and the GGA (dot-dashed line). The GL2 
curve (dashed) corresponds to the slope of the exact curve at $\lambda=0$.
Shown is the difference between H$_2$ and two free H atoms,
evaluated on self-consistent EXX densities.
The ``exact'' curve is an interpolation~\cite{com:interpolate} based on
accurate values of $\Delta E\x$, $\Delta E\xc$, and $\Delta U\xc$ from
a configuration interaction calculation~\cite{rvlphd,gri96a}.
}
\label{fig:aceq}
\end{figure}

For the atoms and most chemically bonded systems, the adiabatic connection
curve is rather non-descript, lying close to a straight line.
This is illustrated by H$_2$ at the equilibrium bond length in 
Fig.~\ref{fig:aceq}, where we plot $\Delta U\xc^{\textrm{RPA}}(\lambda)$.
The area enclosed by the adiabatic connection curves represents the 
XC contribution to the dissociation energy, 
$\Delta E\xc$.
As can be seen from Tab.~\ref{tab:aceq} the RPA dissociation energy of
H$_2$ is excellent agreement with the exact value. Noting also the 
good agreement of the endpoints of our RPA curve in Fig.~\ref{fig:aceq}
with accurate data from configuration interaction calculations~\cite{rvlphd,gri96a},
listed in Tab.~\ref{tab:aceq}, this implies that the RPA curve lies very 
close to the true adiabatic connection curve. The straight line 
corresponds to GL2 theory, indicating the $b=1/2$ high density limit
and the initial slope of the true curve given by Eq.~\ref{eq:acgl2}.
For $\lambda > 0$, $\Delta U\xc^{\textrm{GL2}}(\lambda)$ lies below
the true curve, overestimating the absolute correlation energy. 
Indeed, calculating $\chi^{\textrm{RPA+X}}$ to first order in $\lambda$ 
we obtain $\Delta E^{\textrm{GL2}}=-5.04$~eV, about $0.3$~eV below the 
true dissociation energy. This indicates that the second-order 
perturbative treatment is qualitatively but not quantitatively
accurate for H$_2$. Regarding the PBE GGA, Fig.~\ref{fig:aceq} and
Tab.~\ref{tab:aceq} show that it is accurate for the exchange 
energy ($\lambda=0$) but underestimates the absolute correlation 
potential energy ($\lambda=1$). In turn the PBE GGA also understimates
the absolute XC energy and the dissociation energy of H$_2$.
\begin{table}
\caption{Adiabatic decomposition of the dissociation energy $\Delta E$ 
of H$_2$ at bond length $R=1.4$~bohr, evaluated on self-consistent EXX densities.
Shown are the differences between the molecule and two free
H atoms for the coupling strength integrand $\Delta U\xc(\lambda)$
and related quantities, as explained in the text. All values are in eV.
} 
\label{tab:aceq}
\begin{ruledtabular}
\begin{tabular}{lcccccc}
	& $\Delta E$ & $\Delta E\xc$ & $\Delta E\x$ & $\Delta U\c$ & $\frac{\Delta T\c}{|\Delta U\c|}$ & $\Delta U\xc'(0)$ \\ 
\hline
PBE GGA	& $-4.54$    & $-1.91$       & $-1.04$      & $-1.52$          & $0.431$	& $-2.42$\\
RPA     & $-4.73$    & $-2.10$       & $-0.99$      & $-1.95$          & $0.427$	& $-2.97$\\
exact   & $-4.74$\footnotemark[1]
                     & $-2.04$\footnotemark[2]
                                     & $-0.98$\footnotemark[2]
						    & $-1.93$\footnotemark[2]
                                                                       & $0.450$
										        & $-2.84$\\
\end{tabular}
\end{ruledtabular}
\footnotetext[1]{Reference~\cite{kol60a}.}
\footnotetext[2]{From Refs.~\cite{rvlphd,gri96a}. In these works,
$E\x$, $E\xc$, and $T\c$ were calculated on the H$_2$ density
obtained from a configuration interaction calculation which yielded
$\Delta E = -4.68$~eV.
Using these data we evaluated $\Delta U\c$ from Eq.~\ref{eq:tc}.
}
\end{table}

Calculations of the exact adiabatic connection, using the
accurate ground-state wavefunctions for all $\lambda$,
have so far not been attempted to our knowledge for molecules, 
while a few such curves based on accurate ground-state densities
have been reported for atoms~\cite{puz01a,sav01a,sav03a}, bulk Si~\cite{hoo98a}, and model 
systems~\cite{mtb03}. We remark that $\frac{d}{d\lambda}U\xc[\n](\lambda)|_{\lambda=0}$,
i.e. the GL2 correlation energy, is also a key ingredient
in a recent coupling strength interpolation of the adiabatic
connection by Seidl~{\it et al.}~\cite{sei00a} which performs 
with similar accuracy as modern hybrid functionals for molecular 
dissociation energies. For H$_2$ close to the equilibrium bond
length, the corresponding dissociation energy and adiabatic 
connection curve are in good agreement with our RPA results~\cite{com:isi}.
 
\subsection{H$_2$ symmetry dilemma}

\begin{figure}
\includegraphics[clip=,width=0.9\linewidth]{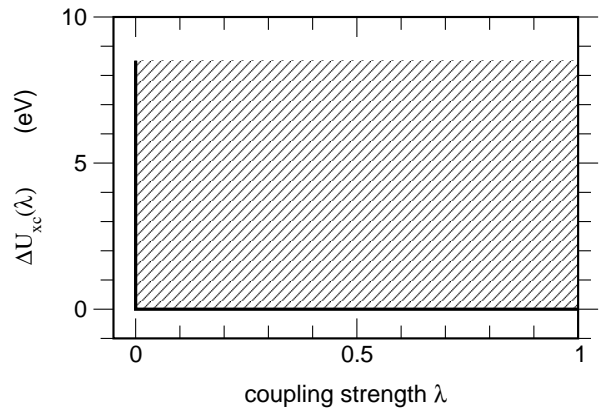}
\caption{Exact adiabatic connection for dissociated H$_2$ at
bond length $R\rightarrow \infty$, shown as the difference
with respect to two free H atoms. For the exact KS determinant
the curve starts at $\lambda=0$ with the negative of the exact
exchange energy of a single H atom, $-E\x(\mbox{H})\simeq 8.5$~eV
as explained in the text. The negative of the shaded area represents the 
correlation energy of the H$_2$ molecule and equals $E\x(\mbox{H})$.
}
\label{fig:acinf}
\end{figure}

We now discuss the stretching of H$_2$ as a paradigm of the
difficulties that single-determinant methods have with 
dissociation. The nearly straight line behavior
dramatically changes 
when the bond is stretched to $R\rightarrow\infty$.
Asymptotically, the proper molecular wavefunction for any $\lambda > 0$ 
(i.e. regardless of the interaction strength) is the bonding
linear combination of the $1s$ orbitals of the two H
atoms. For the H$_2$ ``super-molecule'' the exchange
energy therefore has the same value as in
a hydrogen atom, i.e.
$E\x[\mbox{H$^{\cdot}{\cdots}$\,H$^{\cdot}$}]\rightarrow E\x[\mbox{H}]=-U[\n_{\rm H}]$.
Consequently also $U\c[\mbox{H$^{\cdot}{\cdots}$\,H$^{\cdot}$}](\lambda) \equiv -U[\n_{\rm H}]$ for
any $\lambda>0$, since the dissociated H$_2$ molecule must have 
the same total energy as two H atoms.
Figure~\ref{fig:acinf} shows the
corresponding exact adiabatic connection 
$\Delta U\xc[\mbox{H$^{\cdot}{\cdots}$\,H$^{\cdot}$}](\lambda)$.
The immediate drop of the adiabatic connection curve at $\lambda=0$ is 
characteristic for a system with strong static correlation: here $b=0$ exactly.
The 
position of one electron entirely determines 
the position of the other electron: the two electrons in 
infinitely separated H$_2$ must sit on the two different 
nuclei but never on the same, as spuriously allowed by the single KS
determinant. This type of correlation is often refered to 
as left-right or static correlation, where the true 
many-electron wavefunction takes on multi-determinant 
character~\cite{kh01a}. Put differently~\cite{bae01a}, in
the concept of the XC hole, the exchange hole of H$_2$ is 
spatially completely delocalized over both nuclei. However 
the XC true hole is always centered about the reference 
electron. This means that the correlation hole must be 
long-ranged to yield the proper hydrogen-like hole on one 
nucleus and the needed zero total XC hole on the opposite 
nucleus. By contrast LDA or GGA correlation holes,
derived essentially from the uniform electron gas, are
always short-ranged, and hence cannot cancel the exact
exchange hole far away. Breaking inversion symmetry, L(S)DA 
and spin-dependent GGA on the other hand (like unrestricted 
Hartree-Fock) already yield the spin-up and -down exchange 
holes of separate hydrogen atoms, opposite spin electrons 
sitting on different nuclei, and thus mimic the static 
correlation: unrestricted Hartree-Fock or exact KS exchange 
indeed yield two hydrogen atoms as the dissociation products,
and produce curves similar to Fig.~\ref{fig:acinf}.

\subsection{Behavior of hybrid functionals}

\begin{figure}
\includegraphics[clip=,width=0.9\linewidth]{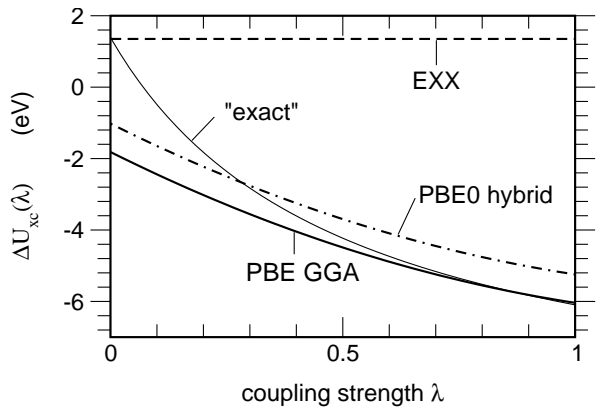}
\caption{Sketch of $\Delta U\xc(\lambda)$ for N$_2\,\rightarrow 2$\,N.
(The EXX, PBE, and PBE0 hybrid curves are calculated 
from the self-consistent PBE densities and orbitals. 
The ``exact'' curve is an interpolation between 
$\Delta E\x$ and $\Delta U\xc^{\textrm{PBE}}$,
following Ref.~\cite{ern96a}.)
}
\label{fig:acn2}
\end{figure}

Does any of this matter for real systems, especially if we
never ask them to dissociate?  The answer is emphatically yes.
Currently our most accurate popular functionals are hybrids,
which mix a fraction $a_0\approx 1/4$ of exact exchange with GGA.
To understand why this improves the accuracy over GGAs~\cite{ern97a}, 
one first notes that usually the latter work best near $\lambda=1$
(H$_2$ being an exception to this rule)~\cite{ern97a}. 
But they worsen as $\lambda\rightarrow 0$, especially when 
there is static correlation in the system. The reason is 
twofold and illustrated in Fig.~\ref{fig:acn2} for N$_2$ (where
static correlation is expected to be most pronounced 
in the $\pi$ bonds). 
On the one hand GGA exchange energies tend to be too negative 
for molecules, leading to a too low starting point for the adiabatic 
connection at $\lambda=0$. On the other hand the GGA correlation 
potential energy is accurate toward $\lambda=1$, but too flat 
a function otherwise. As a consequence $\Delta U\xc(\lambda)$
in the GGA lies too low, leading to an overestimate 
of the magnitude of the dissociation energy. Strong static
correlation shows up as a rapid drop of $U\xc(\lambda)$ 
from $\lambda=0$ toward $\lambda=1$. Note that the free N 
atoms contain little static correlation, that is, their 
$U\xc(\lambda)$ lie close to a straight line and 
can be accurately obtained from GL2 perturbation theory. 
Recognizing that GGAs (and also the LDA) are typically least 
accurate at the $\lambda=0$ end, hybrid functionals can improve 
over the GGA adiabatic connection by mixing in the exact
exchange energy, i.e. part of the exact limit $U\xc(0)=E\x$.
This has led to hybrid XC functionals of the generic form 
\begin{equation} 
\label{eq:hybrid}
E\xc^{\rm hyb}=E\xc^{\rm GGA}+a_0(E\x-E\x^{\rm GGA})
\quad.
\end{equation}
Becke has determined an emprirical value of $a_0\approx 0.28$ ~\cite{bec93a}. 
The non-empirical estimate of $a_0=1/4$ has been given, based on the observation 
that $U\xc(\lambda)$ is formally provided by DFT perturbation theory 
and fourth order is adequate for typical molecules, and defines the so-called 
PBE0 hybrid~\cite{per96a}, shown for N$_2$ in Fig.~\ref{fig:acn2}. 
Hybrid functionals like the PBE0 on the average yield molecular
dissociation energies close to chemical accuracy~\cite{ada99a}, 
improving over GGA.
Pictorially, the admixing of exact exchange shifts the GGA curve a
fraction $a_0$ toward the true starting point ($\lambda=0$) of the correct
curve. Since the true adiabatic connection curve drops sharply from 
$\lambda=0$, a fraction $a_0<1/2$ produces a hybrid curve whose
area is about that of the true curve, as in Fig.~\ref{fig:acn2}.

{}From the adiabatic connection we can also understand when 
(semi-)local and hybrid functionals become inappropriate. 
If the XC energy is dominated by exchange and there is
little correlation energy, the adiabatic connection curve 
drops little from its $\lambda=0$ value. For one-electron 
systems, such as in H$_2^+$, there is even zero correlation 
and the adiabatic connection curve stays completely flat. 
Then one ought to use $a_0\approx 1$, i.e. 100\% exact 
exchange. Indeed GGA as well as hybrid functionals perform 
poorly for one-electron bonds such as in radical ions~\cite{sod99a,bal97a}. 
On the other hand if static correlation is important, the
adiabatic connection will drop rapidly from its value at 
$\lambda=0$, given by the exact exchange energy.
This drop is missed by GGA correlation functionals, which 
then underestimate the correlation energy. This error may be 
compensated by using approximate GGA exchange 
functionals since these tend to overestimate the absolute exchange
energy in molecules and the exchange contribution to 
dissociation energies~\cite{ern97a}. 
Hybrids with $a_0 > 0$ improve the exchange energies over
GGA alone but eventually also reduce the error cancellation due to GGA 
exchange. An early indication for this was the observation
that combining 100\% exact exchange with GGA correlation 
completely fails for molecular dissociation energies~\cite{cle90a}. 
In fact no exact exchange mixing ($a_0\approx0$) can sometimes 
be the better choice as demonstrated in Ref.~\cite{ern97b}. 
We stress that to describe strong static correlation effects 
in bond dissociation, any of these functionals must be used
in spin-unrestricted form in order to attain exchange energies 
that are negative enough to compensate for the lack in the 
respective approximation for correlation.
Clearly, to describe the transition from
exchange dominated to static correlation dominated XC regimes, 
a more flexible, i.e. ``system sensitive'' interpolation 
of the adiabatic connection than provided by present hybrid 
functionals is needed~\cite{BEP97}.

\subsection{Beyond the Coulson-Fisher point}
\label{sec:dissociate}

A key concept in this paper is that DFT within the RPA allows 
correct dissociation of molecules. However, given our present
inability to perform self-consistent RPA calculations, the 
demonstration of this fact becomes quite subtle. While the
(restricted) EXX solution is an adequate starting point for
the RPA around the equilibrium
bond length, it is totally inadequate beyond the Coulson-Fischer
point ($R > 2.5$~bohr for H$_2$ treated in EXX), where ambiguity arises in 
a single-determinant calculation.
In the words of the symmetry dilemma, should one use the unrestricted
solution, which has a pretty good energy but totally incorrect
spin density, or the restricted solution, which has the correct
symmetry but poor energetics (as seen in Fig.~\ref{fig:pes})~?

The answer has been given in Ref.~\cite{per95a}. One must use the 
best estimate for the correct ground-state density 
that is available. As argued there, the unrestricted solution yields the 
best approximate DFT density, but its spin density is not to be 
believed. We therefore take the {\em total} density from our 
{\em unrestricted} EXX KS calculation, and treat it as a spin-singlet. 
This becomes our input density to our RPA calculation. Recall
that this density becomes exact in the limit of $R\rightarrow\infty$, 
where it corresponds to two separate hydrogenic densities, in contrast 
to a {\em restricted} scheme. Inverting the KS equation~\cite{umr94a} 
for $n(\vec r) = n_{\uparrow}^{\textrm{EXX}}(\vec r) + n_{\downarrow}^{\textrm{EXX}}(\vec r)$, we 
obtain the KS potential yielding that density and the respective 
KS eigenstates.

\section{H$_2$ dissociation within the RPA}
\label{sec:results}

\begin{figure}
\includegraphics[clip=,width=0.9\linewidth]{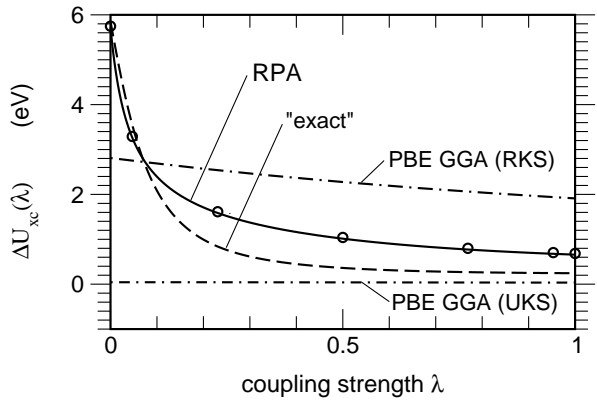}
\caption{Same as Fig.~\ref{fig:aceq}, but for $R=5$~bohr, i.e. beyond
the Coulson-Fisher point. The RPA results are based on the {\em total} 
density of a unrestricted EXX KS calculation. Also shown are the adiabatic 
connections for the PBE GGA applied in the restricted KS formalism (RKS), yielding 
poor energetics, and in the unrestricted KS formalism (UKS), yielding better energetics
but artifically breaking inversion symmetry. 
}
\label{fig:ac5}
\end{figure}

As can be seen from Fig.~\ref{fig:ac5}, the RPA adiabatic connection 
of H$_2$ beyond the Coulson-Fisher point becomes strongly bent downward, 
capturing the onsetting crucial contribution of static correlation
related to the multi-determinant nature of the interacting many-electron 
wave function. The latter is not captured in the PBE GGA correlation functional 
which significantly underestimates the magnitude of the correlation energy,
as seen from Tab.~\ref{tab:ac5} and evident from the PBE GGA (RKS) curve
in Fig.~\ref{fig:ac5}. 
Switching to the {\em unrestricted} 
PBE (and PBE0) schemes, their exchange component simulates the missing 
static correlation while the correlation component is much too small in
magnitude as Tab.~\ref{tab:ac5} shows.
\begin{table}
\caption{Adiabatic decomposition of the dissociation energy $\Delta E$
of H$_2$ at bond length $R=5$~bohr, as shown in Fig.~\ref{fig:ac5}. 
The RPA functional is evaluated on the total density from a spin-unrestricted 
EXX calculation as described in Sec.~\ref{sec:dissociate}. Also
shown are results for the PBE GGA, evaluated as a spin-restricted
(RKS) and a spin-unrestricted (UKS) functional using the EXX
spin-densities. All values are in eV.
}
\label{tab:ac5}
\begin{ruledtabular}
\begin{tabular}{lcccccc}
        & $\Delta E$ & $\Delta E\xc$ & $\Delta E\x$ & $\Delta U\c$ & $\frac{\Delta T\c}{|\Delta U\c|}$ & $\Delta U\xc'(0)$ \\
\hline
PBE (RKS) & $ 2.20$  & $2.30$        & $2.81$      & $-0.90$       & $0.43$   & $-1.23$\\
PBE (UKS) & $-0.06$  & $0.040$       & $0.044$     & $-0.008$      & $0.50$   & $ 0.084$\\
RPA     & $0.54$    & $1.33$        & $5.72$       & $-5.06$       & $0.13$   & $-111$ \\
exact   & $-0.10$\footnotemark[1]
                    & $0.82$\footnotemark[2]
                                     & $5.85$\footnotemark[2]
                                                    & $-5.61$\footnotemark[2]
                                                                       & $0.10$
                                                                                & $-56.8$ \\
\end{tabular}
\end{ruledtabular}
\footnotetext[1]{Reference~\cite{kol60a}.}
\footnotetext[2]{From Refs.~\cite{rvlphd,gri96a}, see also Tab.~\ref{tab:aceq}.
}
\end{table}
Correspondingly, {\em unrestricted} PBE (like PBE0) eventually
give dissociation energies that are in quite good agreement with 
the exact value, as listed in Tab.~\ref{tab:pes}. We stress 
that this is a result of error cancellation between (unrestricted) 
exchange and correlation: the PBE adiabatic connection 
curves is qualitatively clearly wrong, especially at $\lambda=0$,
where its $\Delta E\x$ is much too small and its slope turns out
even slightly positive (though this is not visible on the scale 
of Fig.~\ref{fig:ac5}).
No such error cancellation occurs within the RPA. 
The onset of strong static correlation is reflected in the low value
of our correlation strength parameter $b$ reported in Tab.~\ref{tab:ac5}
for the RPA and exact curves, but not for (semi-) local density 
functional approximations. 

\begin{figure}
\includegraphics[clip=,width=0.9\linewidth]{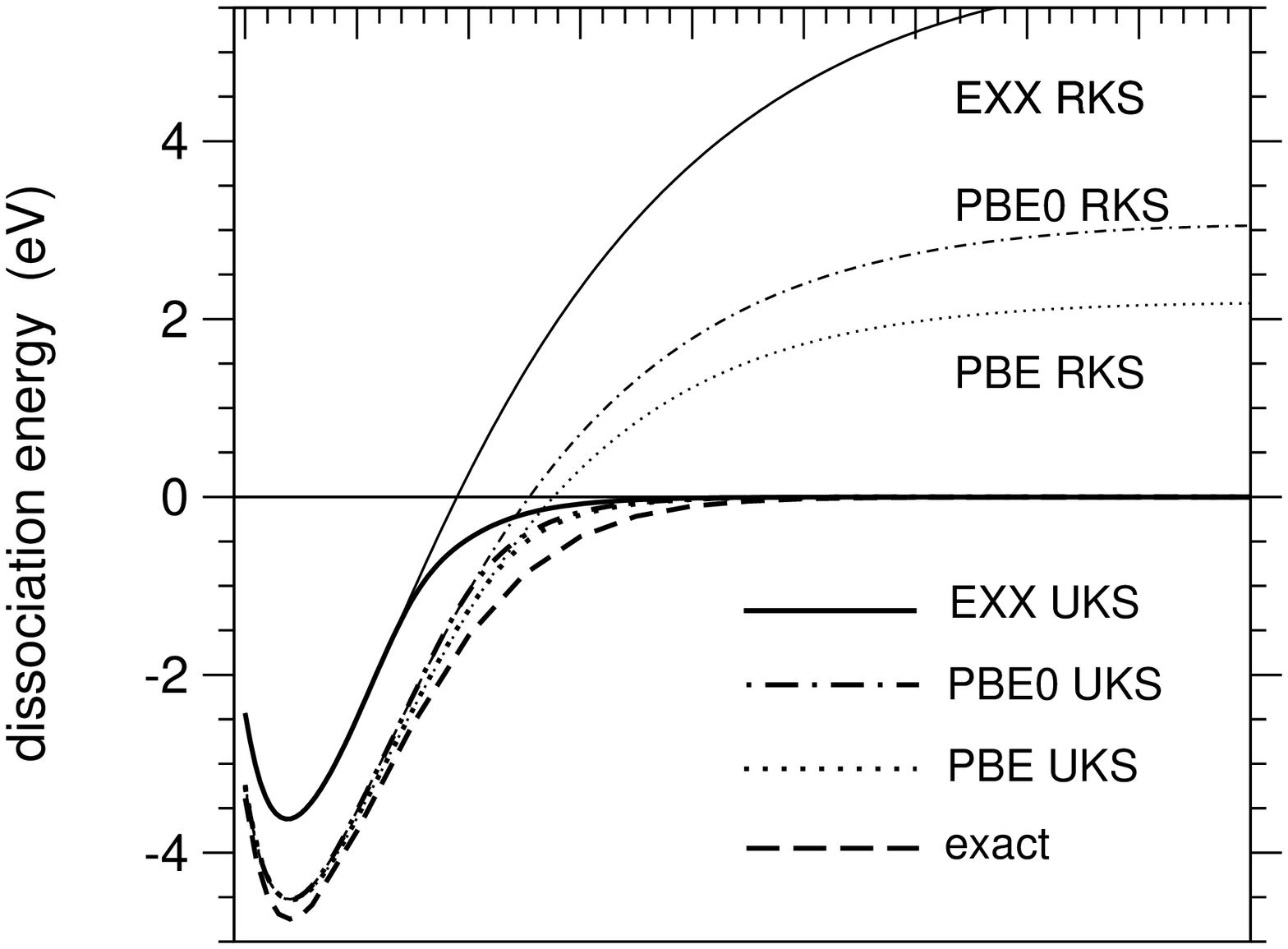}
\\[+1ex]
\includegraphics[clip=,width=0.9\linewidth]{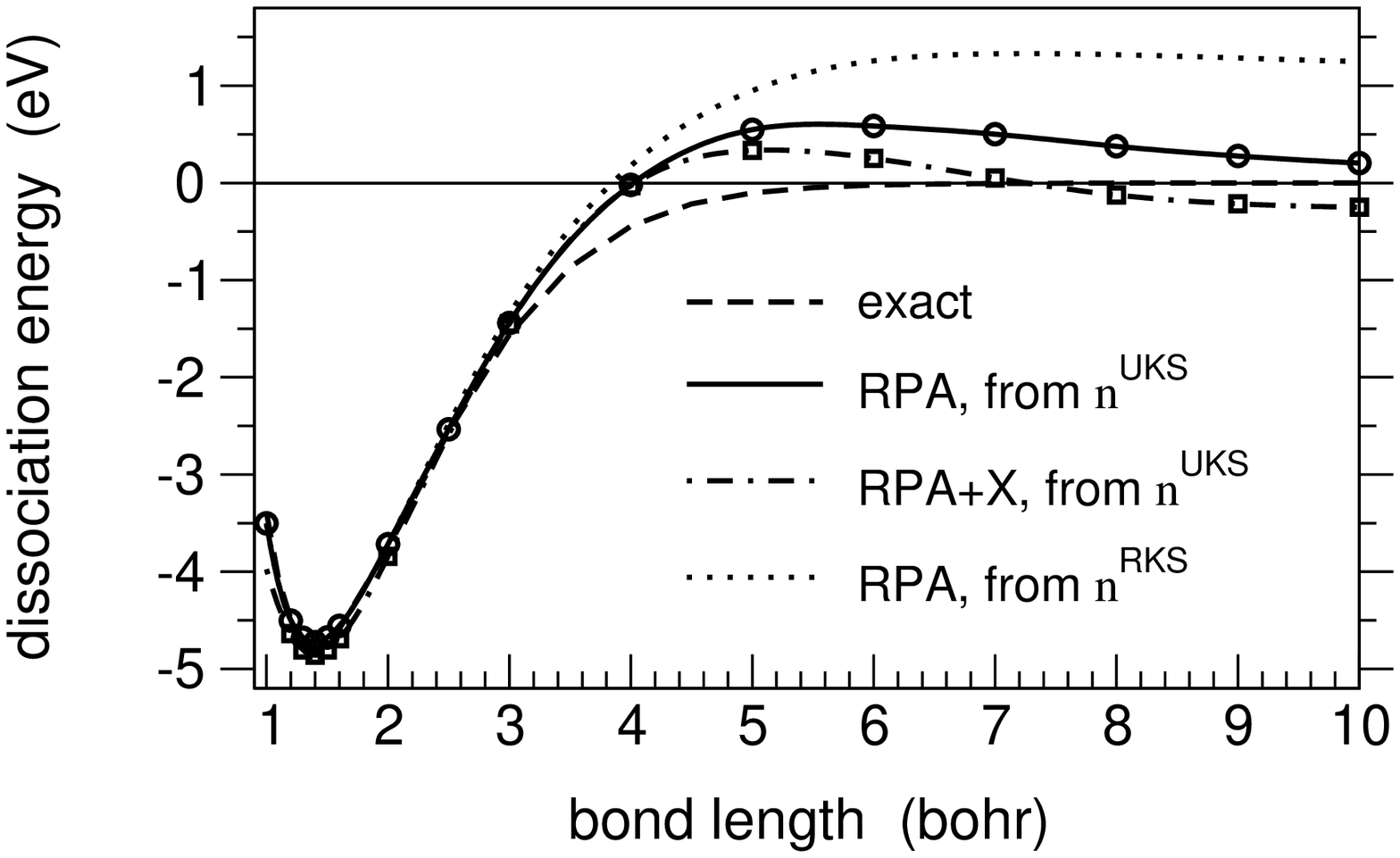}
\caption{Dissociation energy curve of H$_2$. The upper panel compares results 
for EXX, the PBE GGA, and the PBE0 hybrid functionals calculated in the
self-consistent restricted (RKS) and unrestricted (UKS) KS formalism,
with exact data from Ref.~\cite{kol60a}.
The lower panel compares the RPA and RPA+X curves, calculated with {\rm total} 
densities $n^{\rm UKS}$ obtained from unrestricted EXX KS calculations,
with exact data. Also shown is the RPA curve calculated with densities $n^{\rm RKS}$ from
restricted EXX KS calculations.
}
\label{fig:pes}
\end{figure}

Although qualitatively correct the RPA adiabatic connection curve is
still deficient, as can be appreciated by comparing to the exact
curve in Fig.~\ref{fig:ac5} and data in Tab.~\ref{tab:ac5}. $\Delta U\c^{\textrm{RPA}}(\lambda)$
does not drop deep enough with $\lambda$, despite its too steep initial slope.
Correspondingly the RPA yields a too positive molecular correlation energy
and produces an artificial barrier for dissociation as seen in Tab.~\ref{tab:pes}.
\begin{table}[b]
\caption{Dissociation energy of H$_2$ at bond length $R$,
calculated with different XC functionals as indicated in
Fig.~\ref{fig:pes}. Given are the energies from unrestricted
KS calculations, except for RPA and RPA+X as explained
in the text. All values are in eV.
}
\label{tab:pes}
\begin{ruledtabular}
\begin{tabular}{lcccr}
$R$ (bohr)      & $1.4$         & $3$           & $5$           & $10$    \\
\hline
EXX             & $-3.62$       & $-0.47$       & $-0.02$       & $0.00$  \\
PBE GGA         & $-4.53$       & $-1.27$       & $-0.03$       & $0.00$  \\
PBE0 hybrid     & $-4.52$       & $-1.07$       & $-0.03$       & $0.00$  \\
RPA             & $-4.73$       & $-1.44$       & $+0.54$       & $+0.20$ \\
RPA+X	        & $-4.86$       & $-1.45$       & $+0.34$       & $-0.25$ \\
\hline
exact\footnotemark[1]
                & $-4.75$       & $-1.56$       & $-0.10$       & $0.00$
\end{tabular}
\end{ruledtabular}
\footnotetext[1]{Reference~\cite{kol60a}.}
\end{table}
This is further evidenced in the full RPA dissociation 
curve of Fig.~\ref{fig:pes}. Indeed, while the RPA performs accurately around 
equilibrium $R$ and again at larger $R$, it shows an unphysical bump 
at intermediate bond length $R$. The origin of this bump will be further
discussed below. Nonetheless the asymptotic behavior ($R\rightarrow\infty$) 
of the RPA is correct, as can be
understood from a model RPA calculation using only the HOMO and LUMO EXX-KS states.
As shown in the Appendix, this model RPA calculation  yields the exact correlation (and total) 
energy for $R\rightarrow\infty$, and produces precisely the exact 
adiabatic connection of Fig.~\ref{fig:acinf}.
Including the higher lying KS states, the RPA builds up spurious 
self-correlation in both the H atoms and the H$_2$ super-molecule,
which however cancels out in the dissociation energy. This cancellation 
is indeed also reflected in the (identical) estimates of the 
local-density corrections (RPA+) in the atom and in the infinitely 
stretched molecule.

Figure~\ref{fig:pes} also demonstrates that for proper dissociation 
it is crucial to work with qualitatively correct densities, 
i.e., to start from a KS potential that takes into account the
essential left-right correlation. 
Describing the system by the restricted KS formalism using the 
EXX, PBE GGA or PBE0 hybrid functionals leads to much too high 
total energies for the dissociating bond because the self-consistent
EXX density is {\em qualitatively} wrong as explained in 
Sec.~\ref{sec:dilemma}. While the PBE GGA and PBE0 include correlation, 
improving over EXX, it is obvious from Fig.~\ref{fig:pes} that
only the unrestricted KS formalism leads to reasonably accurate
dissociation energies for larger $R$. The fact that the PBE0
curve raises too quickly above the true curve indicates that
the error cancellation between exchange and correlation is
significantly incomplete, as explained in Sec.~\ref{sec:ac}. 
For H$_2$, PBE GGA clearly works better, and the lower energy
solution appears only beyond $R\approx 4$~bohr, i.e. for
a markedly larger bond length than in both the EXX and the
PBE0 hybrid. In Table~\ref{tab:pes} we list the 
dissociation energies for the different functionals at 
various bond lengths.

The RPA curve is accurate around the H$_2$ equilibrium bond 
length and approaches the dissociation limit for large $R$. 
The success of the RPA lies in the fact that it does so (properly) as a 
functional of a singlet density only, rather than 
spin densities as in the traditional approximations for XC. 
Of course the density must ultimately come from a self-consistent 
KS calculation and potential, whereas we have approximated it 
non-selfconsistently. Our findings confirm that proper densities 
require accurate approximations also for the XC potential (the 
functional derivative of $E\xc[n]$), as argued recently by 
Baerends~\cite{bae01a}. In agreement with Ref.~\cite{bae01a}, our results also 
show that unoccupied KS states (and in particular the LUMO state) 
must be included in $E\xc[n]$ in order to attain the correct 
dissociation limit.

An obvious shortcoming of the RPA compared to the unrestricted 
PBE GGA and PBE0 schemes is that its dissociation curve 
displays an unphysical bump for intermediate $R$ as seen in
Fig.~\ref{fig:pes} and Tab.~\ref{tab:pes}. A similar behavior
has been found for N$_2$~\cite{fur01a} and Be$_2$~\cite{fuc02a,fuc02b}.
We believe that these deficiencies stem from the RPA itself, rather 
than our lack of self-consistency. In particular we have obtained
essentially the same curves when we used densities derived from
unrestricted LDA and GGA calculations. A more conclusive answer
must either start from more accurate densities or await self-consistent
calculations. Using an ansatz for the XC energy functional in terms of 
natural orbitals in a (approximately self-consistent) {\em singlet} KS 
calculation, Baerends and coworkers~\cite{bae01a,gru03a} recently 
reported a dissociation curve of H$_2$ in very good agreement with 
the true curve for all $R$. The fact that for intermediate $R$ their 
curve is distinctly more accurate than our non-self-consistent RPA 
result clearly calls for further analysis of both approaches, 
including possible error cancellation between different components 
of the total energy. This is beyond the scope of our present study.

\section{Extensions beyond the RPA}
\label{sec:extensions}

So far we have shown that the RPA gives a qualitatively correct
account of the (differential) adiabatic connection of H$_2$, 
in contrast to semilocal or hybrid functionals, but needs to 
be improved for moderately large bond lengths. We now discuss
possible extensions of the RPA as an ACFDT XC functional. 

A known deficiency of the RPA is the spurious self-correlation 
in the absolute correlation energies, for the H atom as well as 
the H$_2$ molecule. For the one-electron H atom, self-correlation is eliminated
by including the exact exchange kernel ($f\x^{\lambda}
= -\lambda v_{\rm ee}$) in the screening of the electronic Coulomb 
interaction, Eq.~\ref{eq:dyson}. For the two-electron H$_2$ the exact
exchange kernel ($f\x^{\lambda}= -\frac{\lambda}{2}v_{\rm ee}$~\cite{pet96a}), 
eliminates self-correlation to second order in $\lambda$ yet not to higher order. 
Using this RPA+X functional
we obtain the dissociation energy of H$_2$ ($R=1.4$~bohr) as $-4.86$~eV,
about $0.1$~eV below the true and our RPA values. Having estimated
our computational accuracy at the same order, we feel cautious 
about the significance of the RPA+X result. On the other hand
we find that the dissociation energy curve beyond the Coulson-Fisher 
point is not at all improved. To the contrary, while the RPA+X 
curve follows the RPA curve up to $\approx 4$~bohr, it drops below 
zero and approaches a negative constant for larger $R$. We regard 
this as a size-consistency problem in that the self-correlation error
is eliminated in the H atoms, but reappears in the far stretched H$_2$. 
This is further corroborated in Appendix B. Our finding 
clearly suggests that in addition to exchange also correlation 
contributions need to be included in the XC kernel. A ready way to 
do so would be to employ the well-known adiabatic LDA kernel~\cite{lei99a} 
or an energy-optimized adiabatic XC kernel~\cite{dob00b}.

As mentioned in Sec.~\ref{sec:rpa} the incorrect RPA short-ranged
correlation, including self-correlation, may be corrected through
a specially designed (semi-)local-density functional (RPA+). Although 
this improves upon the too negative RPA correlation energies
as shown in Ref.~\cite{fuc02a}, it again does not correct the
deficiencies we observe in the RPA dissociation energy curve of
H$_2$.

The limitation of RPA, as in any adiabatic treatment of the interacting
linear response, might be that it treats only single excitations~\cite{cas96a} 
and thus cannot take into account contributions to density fluctuations 
that correspond to doubly excited determinants and eventually also contribute 
to the correlation energy. 
As is well known, doubly excited determinants have larger weights in the 
asymptotic {\em interacting} wavefunctions of H$_2$ for both ground and 
excited states. For $R\rightarrow\infty$ this is evident from the corresponding 
exact Heitler-London wavefunctions (see e.g. Refs.~\cite{slater} 
and \cite{gri00a}; for instance, the lowest $1\Sigma_g^+$ and $1\Sigma_u^+$ singlets
are the symmetric and antisymmetric linear combinations, respectively, of 
the determinants $|\sigma_g\overline\sigma_g|$ and $|\sigma_u\overline\sigma_u|$ 
made up of the HOMO and LUMO). 
Double excitations imply additional poles in the (real frequency)
interacting response and thus a strongly frequency-dependent XC kernel.
Any such pole contributes to the spectral decomposition of the 
pair-correlation function or XC hole.
While there is limited progress concerning the calculation of the excitation 
energies for certain doubly excited states within TDDFT~\cite{mzcb04},
further analysis (and development) of the spatial and frequency dependence 
of such XC kernels is needed for applications in ACDFT XC functionals.

\section{Summary}
\label{sec:summary}

The central message of this paper is that the
Random Phase Approximation (RPA) resolves the long-standing
symmetry dilemma encountered in {\em approximate} density functional theory
when breaking the H$_2$ electron pair bond.
We showed that the RPA produces the correct
dissociation limit from a proper singlet KS density, without
the need for artifical symmetry-breaking as in unrestricted
Kohn-Sham theory for traditional local- or generalized-gradient
functionals and hybrid exchange-correlation (XC) functionals.
By analyzing the adiabatic connection, we showed that the RPA
captures correctly the strong static (left-right)
correlation that arises
when the pair bond breaks.
Local and gradient-corrected functionals make serious errors here,
and even hybrids, which mimic this effect at equilibrium bond lengths,
cannot account for the extreme static correlation in the dissociation
limit. 
As the RPA yields an orbital-dependent XC functional, our results
demonstrate the importance of including unoccupied KS states,
in particular the LUMO state.
We also showed that it is crucial to work with an accurate
density, which we have
constructed approximately in this study, and that could be
improved by applying the RPA self-consistently. We further
found the RPA dissociation curve to agree well with exact data
near the equilibrium bond length of H$_2$. When the bond is
stretched, it tends to the correct limit, unlike all the common 
restricted Kohn-Sham approaches.
Noting that the RPA still leads to an
unphysical repulsion of the hydrogens for intermediate bond lengths,
we adressed inherent limitations and possible extensions of the RPA.
Seen as a first step to realize fully nonlocal XC
functionals by the adiabatic-connection fluctuation-dissipation
formalism, we believe that the RPA provides a sound basis for
quantitative refinements. Our study highlights H$_2$ as a
significant benchmark system for assessing future progress
beyond the RPA.
\\

\section*{Acknowledgments}

We thank Ulf von Barth and Michael Seidl for discussions, and
Robert van Leeuwen for comments on the Appendix and kindly 
providing us with his data of Ref.~\cite{rvlphd}.
X.G. thanks the FNRS (Belgium) for financial support.
We acknowledge financial support from the Communaut\'e Francaise de Belgique (through
a subvention ``Actions de Recherche Concert\'ees"), the Belgian Federal State (``Poles
d'attraction Interuniversitaires", Phase V),
and the European Union (contracts HPRN-CT-2002-00317, ``EXCITING" Research Training Network
``First-principles approach to the calculation of optical properties of solids",
and NMP4-CT-2004-500198, ``NANOQUANTA" Network of Excellence
``Nanoscale Quantum Simulations for Nanostructures and Advanced Materials" ).
K.B. was supported by the National Science Foundation under grant CHE-0355405.
Some of this work (K.B.) was performed at the Centre for Research in Adaptive
Nanosystems (CRANN) supported by the Science Foundation Ireland (Award 5AA/G20041).

\section*{Appendix}

In part A of this Appendix we describe our computational
method for evaluating the RPA XC energy.
In part B we corroborate that the RPA total
energy of H$_2$ at asymptotically large bond distances 
from a {\em spin-restricted} Kohn-Sham groundstate 
(i) becomes indeed equivalent to the total energy of two free 
H atoms as calculated in a {\em spin-unrestricted} formalism, 
in contrast to the case of the RPA+X kernel, (ii) 
for sufficiently large separations $R$ includes the 
expected $-C_6^{\rm RPA}/R^6$ van der Waals attraction.

\subsection{Computational Method}
\label{sec:method}

We have implemented the RPA functional in a pseudopotential 
plane-wave framework~\cite{abinit}, handling the response functions, 
Eqs.~\ref{eq:chiks} and \ref{eq:dyson}, in their reciprocal 
space representation. Gauss-Legendre quadrature rules are 
used for the $\lambda$ and $iu$ integrations. For our study
of H$_2$ the $-1/r$ attraction is replaced by a highly accurate 
norm-conserving pseudopotential~\cite{fuc99a}. The latter yields 
practically the exact energy of the H atom and dissociation 
energies to within $0.1$~mHa when compared to full-potential results, for
LDA, GGA, or EXX calculations. We place the H$_2$ molecule
in a fcc supercell of $21$~bohr side length. For the initial
KS calculation we use a plane-wave cutoff energy of $30$~Ha,
and $12$~Ha for the response functions. In the KS response
we include unoccupied states up to $2.5$~Ha explicitly 
and treat the higher ones through a closure relation.
For the frequency integration we employ 12 supports for
$R<2$~bohr and up to 54 for larger $R$, concomitant
with the closing of the HOMO-LUMO gap. For the coupling strength
integration we use 4 to 11 supports to capture the stronger
curvature of $U\xc^{\lambda}$ with increasing $R$. 
From convergence tests we estimate that our total and dissociation
energies are converged to well within $0.1$~eV. Indeed our
value for the H$_2$ dissociation energy in RPA, $-4.73$~eV, 
is in excellent agreement with previous work~\cite{fuc02a,fur01a}.

\subsection{RPA XC for H$_2$ at large bond lengths}
\label{sec:appB}

In this section we show analytically that within the RPA
$\lim_{R\rightarrow\infty} E_{\rm tot}(\mbox{H}_2) = 2 E_{\rm tot}(\mbox{H})$,
i.e. that the total energies of the infinitely stretched, {\em spin-compensated}
H$_2$ molecule and of two separate {\em spin-polarized} H atoms are 
identical. 
We also show that this does not
hold for the exact exchange kernel (RPA+X approximation).
We will examine the density response of the H$_2$ molecule using the particle-hole 
formulation (in a product basis of the KS states) of time-dependent 
DFT~\cite{cas96a}, which is equivalent to the matrix formalism
used in Sec.~\ref{sec:rpa} but more convenient for formal
analysis. We emphasize that our analysis holds for the RPA as 
an adiabatic approximation only, and does not address the effects 
of double excitations that require a non-adiabatic (frequency-dependent)
correlation kernel as argued in Sec.~\ref{sec:extensions}.

For the spin-compensated molecule the response function
for some coupling strength $\lambda$ can be written as
\begin{equation}
\chi^{\lambda}[\mbox{H}_2](\vec r,\vec r';\omega)
= \sum_n Q^{\lambda}_n(\vec r) \frac{4\Omega_n(\lambda)}{\omega^2 - \Omega_n^2(\lambda)}{Q^{\lambda}_n}^{\ast}(\vec r')
\quad,
\end{equation}
where $\Omega_n(\lambda)$ are the transition frequencies and $Q^{\lambda}_n$ are the associated 
amplitudes. 
The $\Omega_n(\lambda)$ are the positive square roots of the eigenvalues of the matrix
\begin{equation}
\label{eq:casida}
M_{ij,kl}(\lambda)=\omega_{ij}^2 \delta_{ik} \delta_{jl}
                       + 4\sqrt{\omega_{ij}}f^{{\sss HXC}}_{ij,kl}(\lambda)\sqrt{\omega_{kl}}
\quad
\end{equation}
involving all possible KS excitations from an occupied KS state $\phi_i$ to an unoccupied KS 
state $\phi_j$, with respective KS transition frequencies $\omega_{ij}=\epsilon_j - \epsilon_i$. Here 
$f^{{\sss HXC}}_{ij,kl}(\lambda)=\int d^3r d^3r' \phi_i(\vec r)\phi_j^{\ast}(\vec r)
f_{{\sss HXC}}^{\lambda}(\vec r,\vec r') \phi_k^{\ast}(\vec r')\phi_l(\vec r')$ 
denotes the matrix elements of the Hartree and XC kernel, and it is assumed that 
one works in the adiabatic approximation, i.e. that the XC kernel is frequency 
independent. From the eigenvectors $U^{\lambda}_{ij,n}$ of $M_{ij,kl}(\lambda)$, 
one obtains the spectral components 
\begin{equation}
Q^{\lambda}_{ij,n}=\sqrt{\omega_{ij}/\Omega_n(\lambda)}\,\,\, U^{\lambda}_{ij,n}
\end{equation}
of the amplitudes 
$
Q^{\lambda}_n(\vec r) = \sum_{i}^{\textrm{occ.}}\sum_j^{\textrm{unocc.}} Q^{\lambda}_{ij,n}\phi_{i}^{\ast}(\vec r)\phi_j(\vec r)
$.
In the RPA, $f_{{\sss HXC}}^{\lambda}=\lambda v_{\textrm{ee}}$.

Consider now the H$_2$ molecule at large $R$. For any finite number 
$N$ of (bound) KS states, $R$ can be chosen large enough such that 
the molecular orbitals can be approximated by the bonding and antibonding 
linear combinations of the atomic orbitals $a_i(\vec r)$ and $b_i(\vec r)$
of the H atoms A and B respectively:
\def\Sn{{S_{\mp}}}
\begin{equation}
\phi_{i\pm}(\vec r) \simeq \left(2\mp 2S_i\right)^{-1/2}\left\{a_i(\vec r) \pm b_i(\vec r)\right\}
\quad,
\end{equation}
where $S_i$ is the overlap integral.
Similarly, we approximate the respective eigenenergies 
$E_{i\pm}\simeq \epsilon_{i}$ by the atomic eigenenergies 
$\epsilon_{i}$, except for the HOMO and LUMO energies $E_{0\pm}$
whose gap we write as $E_g = E_{0-} - E_{0+}$.
 
In the (imaginary frequency) KS response $\chi^0[\mbox{H}_2]$ of the molecule, we 
first split off the HOMO-LUMO transition which is well separated
from the higher ones, and define:
\begin{equation}
\label{eq:delchiks}
\delta\chi^{0}[\mbox{H}_2](\vec r,\vec r',iu)=-4 E_g\frac{q(\vec r)q(\vec r')}{u^2 + E_g^2}
\quad,
\end{equation}
where $q(\vec r)=\phi_{0+}(\vec r)\phi_{0-}(\vec r)=(2\sqrt{1-S_0^2})^{-1}(|a_0(\vec r)|^2-|b_0(\vec r)|^2)$. For $R\rightarrow\infty$,
the remainder $\tilde\chi^0[\mbox{H}_2]=:\chi^0[\mbox{H}_2]-\delta\chi^0[\mbox{H}_2]$ can readily be shown to 
split (up to exponentially decreasing corrections) into atomic contributions $\chi^0[\mbox{H}_{\textrm A}]$ and $\chi^0[\mbox{H}_{\textrm B}]$, where:
\begin{equation}
\chi^0[\mbox{H}_{\textrm A}](\vec r,\vec r';iu)=-\sum_{j=1}^{N/2} 4(\epsilon_j-\epsilon_0)
	\frac{a_0^{\ast}(\vec r)a_j(\vec r)a_j^{\ast}(\vec r')a_0(\vec r')}{u^2+(\epsilon_j-\epsilon_0)^2}\quad,
\quad
\end{equation}
with a similar definition for $\chi^0[\mbox{H}_{\textrm B}]$. $\chi^0[\mbox{H}_{\textrm A}]$ and $\chi^0[\mbox{H}_{\textrm B}]$ are formally equivalent to the KS reponse of a free 
{\em spin-polarized} H atom (see however the discussion of RPA+X below). Hence we asymptotically get:
\begin{equation}
\label{eq:chiksh2}
\chi^0[\mbox{H}_2]
= 
\chi^0[\mbox{H}_{\textrm A}] + \chi^0[\mbox{H}_{\textrm B}] + \delta\chi^0[\mbox{H}_2] + {\cal O}(\mbox{exp.})
\quad.
\end{equation}

As for the asymptotic RPA response ($R\rightarrow\infty$), inspection of Eq.~\ref{eq:casida} 
shows that the lowest eigenvalue $\Omega_0^2(\lambda)$ (i.e. the singlet excitation energy) exponentially tends to zero and is well separated from the others. Indeed,
\begin{equation}
\label{eq:omegazero}
\Omega_0^2(\lambda) = E_g^2 + 4\lambda K_0 E_g + \lambda^2 P(\lambda) E_g +  {\cal O}\left\{(\lambda E_g)^2\right\}
\quad,
\end{equation}
where the HOMO-LUMO exchange integral 
$K_0=\langle\phi_{0-}\phi_{0+}|\hat v_{\textrm{ee}}|\phi_{0+}\phi_{0-}\rangle
\simeq U[\mbox{H}] - (2R)^{-1} + {\cal O}(\mbox{exp.})$
reduces asymptotically to the atomic Hartree energy, and $P$ is an (here) unspecified, 
yet smooth function of $\lambda$. Note that the first two terms on the R.H.S. of 
Eq.~\ref{eq:omegazero} also follow from the single pole approximation 
to TDDFT excitation energies, and that the remainder describes 
corrections due to the coupling with higher KS excitations~\cite{AGB03}. The corresponding eigenvector 
is $U^{\lambda}_{ij,0}=\delta_{ij,0+0-}+{\cal O}(\lambda \sqrt{E_g})$. Hence we can split the asymptotic RPA response of H$_2$ as follows: 
\begin{equation}
\label{eq:chih2}
\chi^{\lambda}[\mbox{H}_2]=\tilde\chi^{\lambda}[\mbox{H}_2]
		+\delta\chi^{\lambda}[\mbox{H}_2] + {\cal O}(\mbox{exp.})
\quad,
\end{equation}
where 
\begin{equation}
\label{eq:delchi}
\delta\chi^{\lambda}[\mbox{H}_2](\vec r,\vec r';iu)=-4 E_g\frac{q(\vec r)q(\vec r')}{u^2+\Omega_0^2(\lambda)}
\quad,
\end{equation}
and
$\tilde\chi^{\lambda}=(1- \tilde\chi^{0}\lambda v_{\textrm{ee}})^{-1}\tilde\chi^{0} $ is the contribution from the other
eigenvalues and eigenvectors of the matrix $M_{ij,kl}(\lambda)$.
For large $R$, we can further write $\tilde\chi^{\lambda}$ as the sum of
the RPA responses of the H atoms, $\chi^{\lambda}[\mbox{H}_{\textrm{A,B}}]$,
and an inter-atomic correction~\cite{dob94a}, $\Delta \chi^{\lambda}[\mbox{H}_2]$\,:
\begin{equation}
\label{eq:chih2tilde}
\tilde\chi^{\lambda}[\mbox{H}_2] = \chi^{\lambda}[\mbox{H}_{\textrm A}]
	+ \chi^{\lambda}[\mbox{H}_{\textrm B}]
	+ \Delta\chi^{\lambda}[\mbox{H}_2]
	\quad .
\end{equation}
The atomic part 
\begin{equation}
\label{eq:chilambdaat}
\chi^{\lambda}[\mbox{H}]=(1- \chi^{0}[\mbox{H}]\lambda v_{\textrm{ee}})^{-1}\chi^{0}[\mbox{H}] 
\quad
\end{equation}
is formally equivalent to the RPA response of a {\em spin-polarized} H atom.

Using the response functions as decomposed in Eqs. \ref{eq:chiksh2}, \ref{eq:chih2} 
and \ref{eq:chih2tilde}, the RPA correlation energy for stretched H$_2$ reads
\begin{widetext}
\begin{eqnarray}
\label{eq:acfdc}
E_{\textrm{c}}^{\textrm{RPA}}[\mbox{H}_2]
	&=& -{\textstyle \int_0^1 d\lambda \int_0^{\infty} \frac{du}{2\pi}}\, \mathrm{Tr}[v_{\textrm{ee}}\{\chi^{\lambda}[\mbox{H}_2](iu)-\chi^0[\mbox{H}_2](iu)\}]\,\,\,
\\[1em]
\label{eq:ech2infty}
      	&=& E_{\textrm{c}}^{\textrm{RPA}}[\mbox{H}_{\textrm A}] + E_{\textrm{c}}^{\textrm{RPA}}[\mbox{H}_{\textrm B}]
		+ \Delta E_{\textrm{c}}^{\textrm{RPA}}[\mbox{H}_2]
		+ \delta E_{\textrm{c}}^{\textrm{RPA}}[\mbox{H}_2]
		\quad .
\end{eqnarray}
\end{widetext}
Here $E\c^{\textrm{RPA}}[\mbox{H}]$ is the RPA correlation energy of a 
{\em spin-polarized} H atom (associated with $\chi^{\lambda}[\mbox{H}]-\chi^0[\mbox{H}]$),
$\Delta E_{\textrm{c}}^{\textrm{RPA}}[\mbox{H}_2]$ comes from $\Delta \chi^{\lambda}[\mbox{H}_2]$, and  
$\delta E_{\textrm{c}}^{\textrm{RPA}}[\mbox{H}_2]$ from $\delta \chi^{\lambda}[\mbox{H}_2]-\delta \chi^{0}[\mbox{H}_2]$.
Expanding $\Delta \chi^{\lambda}[\mbox{H}_2]$
in $\lambda v_{\textrm{ee}}$ one can show, analogously to the case of interacting
closed-shell systems~\cite{dob94a}, that the leading, 
second-order contribution to $\Delta E\c^{\textrm{RPA}}[\mbox{H}_2]$ recovers the van der Waals 
interaction between the H atoms, i.e.
\begin{equation}
\Delta E\c^{\textrm{RPA}}[\mbox{H}_2] \simeq - C_6^{\textrm{RPA}}[\mbox{H}]  R^{-6} \quad ,
\end{equation}
with the atomic $C_6$ coefficient obtained within the RPA. 
We next consider the HOMO-LUMO part of the response, writing $\delta E_{\textrm{c}}^{\textrm{RPA}}[\mbox{H}_2]=\int_0^1 d\lambda\, \delta U\c^{\textrm{RPA}}[\mbox{H}_2](\lambda)$, where
\begin{widetext}
\begin{equation}
\label{eq:deluc}
\begin{array}{rcl}
\delta U\c^{\textrm{RPA}}[\mbox{H}_2](\lambda)
    & = &-\int_0^{\infty}\frac{du}{2\pi}\,
\mathrm{Tr}[v_{\textrm{ee}}\{\delta\chi^{\lambda}(iu)-\delta\chi^0(iu)\}]\\[1em]
	& = & K_0 \left( \frac{E_g}{\Omega_0(\lambda)} - 1 \right)
	\\[1em]
	& = & K_0 \left( \left\{ 1 + 4\lambda\frac{K_0}{E_g}
         + \frac{\lambda^2 P(\lambda)}{E_g} + {\cal O}(\lambda^2) \right\}^{-1/2} - 1\right)
\quad.
\end{array}
\end{equation}
\end{widetext}
The integration over $\lambda$ then yields asymptotically:
\begin{eqnarray}
\delta E\c^{\textrm{RPA}}[\mbox{H}_2] 
&=&
-K_0 + {\cal O}(\sqrt{K_0 E_g})
\label{eq:decrpainfk}
\\
&\simeq&
-U[\mbox{H}] + (2R)^{-1} + {\cal O}(\mbox{exp.})
\quad.
\label{eq:decrpainf}
\end{eqnarray}
The term $\delta E_{\textrm{c}}^{\textrm{RPA}}$ is associated with the static correlation due 
to the (nearly) degenerate HOMO and LUMO KS states. 
Adding 
$E\x[\mbox{H}_2] \sim 2 E\x[\mbox{H}] + U[\mbox{H}] - (2R)^{-1} + {\cal O}(\mbox{exp.})$ 
to
$E\c^{\textrm{RPA}}[\mbox{H}_2]$ 
we last get for the RPA XC energy
\begin{equation}
E\xc^{\textrm{RPA}}[\mbox{H}_2] \simeq 2 E\xc^{\textrm{RPA}}[\mbox{H}] - C_6^{\textrm{RPA}} R^{-6}
\quad .
\label{eq:excrpainf}
\end{equation}
The kinetic, electron-nucleus, nucleus-nucleus and Hartree components 
of the total energy of H$_2$ are easily shown
to also approach those of the free atoms for $R\rightarrow\infty$. Hence the RPA obeys the expected 
result
$\lim_{R\rightarrow\infty} E_{\textrm{tot}}^{\textrm{RPA}}[\mbox{H}_2]=2E_{\textrm{tot}}^{\textrm{RPA}}[\mbox{H}]
$.

Several remarks are in order:\\
{\em Leading $R$-dependence of $E\c^{\textrm{RPA}}[\mbox{H}_2]$ for $R\rightarrow\infty$:}
Equation~\ref{eq:excrpainf} states that the $C_6^{\textrm{RPA}}R^{-6}$ van der Waals term is the leading correction
to the asymtpotic RPA XC energy. This finding rests on the result of Eq.~\ref{eq:decrpainf}
that the static correlation term $\delta E\c^{\textrm{RPA}}[\mbox{H}_2]$ follows the $(2R)^{-1}$
behavior of $K_0(R)$, i.e. that it contains no multipole terms of higher power than $R^{-1}$ up to $R^{-6}$. 
The latter holds if (i) the HOMO and LUMO are represented by linear combinations 
of $s$-like atomic functions, as is appropriate for large $R$ and as we had assumed. Then higher order 
terms in $K_0(R)$ decay in fact exponentially with $R$, as can be seen from a multipole decomposition of $K_0$.
A further condition is that (ii) the HOMO-LUMO gap (and thus $\sqrt{K_0E_g}$ in Eq.~\ref{eq:decrpainfk})
decays exponentially, as is expected for Kohn-Sham states (as opposed to a Hartree-Fock calculation, 
where $E_g\propto R^{-1}$) and which we have verified numerically in the range $R=4\ldots 10$~bohr. 
Of course, for the bond lengths considered here the van der Waals term is in fact marginal:
for instance, $C_6^{\textrm{RPA}}R^{-6}\sim 8$~meV and $\sim 0.1$~meV at $R=5$ and $10$~bohr, 
respectively, i.e. more than an order of magnitude smaller than the total RPA errors given in Tab.~\ref{tab:pes}
(using $C_6^{\textrm{RPA}}\sim 4.6$~a.u., calculated from the atomic dipole polarizability).
Clearly, the RPA dissociation energy curve is still dominated in the range $R=4\ldots 10$~bohr by the 
static correlation term of Eqs.~\ref{eq:decrpainfk} and \ref{eq:decrpainf}, as we further discuss 
in the next paragraph.

{\em Repulsion at intermediate $R$ and role of self-consistency:}
From Eq.~\ref{eq:deluc} we can interpret the ratio
$\alpha^{\textrm{RPA}}(\lambda)=K_0 E_g/\Omega_0^{\textrm{RPA}}(\lambda)<K_0$
as a correction to the exact asymptotic adiabatic connection (Fig.~\ref{fig:acinf})
that decays exponentially with $R\rightarrow\infty$, turning on static correlation. $\alpha^{\textrm{RPA}}(\lambda)$ yields a positive ${\cal O}(\sqrt{K_0 E_g})$ contribution to the RPA exchange-correlation energy (see Eq.~\ref{eq:decrpainf}).
While this contribution is expected to die out exponentially like $E_g$, it may still be significant around $R=10$~bohr, 
showing up as a spuriously repulsive dissociation curve. 
Our $E_g \sim 10^{-2}$~eV at $R=10$~bohr
is indeed compatible with the $\sim 0.2$~eV error we find from our RPA calculation at this bond length (this estimate follows from the single transition model discussed at the end of this appendix).
We cannot rule out that in a self-consistent treatment $E_g$ decays (sufficiently) more rapidly compared
to our present non-selfconsistent calculation.

{\em Behavior of $E\c^{\textrm{RPA+X}}[\mbox{H}_2]$ for $R\rightarrow\infty$:}
For the exact exchange kernel~\cite{pet96a} the same analysis of the molecular
correlation energy
as for the RPA can be carried through with $\lambda$ replaced by $\lambda/2$. For $R\rightarrow\infty$, the static 
correlation term $\delta E\c^{\textrm{RPA+X}}[\mbox{H}_2] \sim -U[\mbox{H}] + (2R)^{-1} + {\cal O}(\mbox{exp.})$
remains the same as in the RPA. However the atomic terms in Eq.~\ref{eq:ech2infty} 
do not vanish, in contrast to the RPA+X correlation energy of {\em spin-polarized} free H atoms.
Indeed, for a free H atom the spin-density responses $\chi^{\lambda\,\textrm{RPA+X}}_{\sigma\sigma'}[\mbox{H}]$ 
and $\chi^{0}_{\sigma\sigma'}[\mbox{H}]$ are identical, as is easily seen from the 
spin-resolved Dyson equation (see e.g.  Ref.~\cite{lei99a}). Thus 
$E\c^{\textrm{RPA+X}}[\mbox{H}] = 0$ and $E_{\textrm{tot}}^{\textrm{RPA+X}}[\mbox{H}]=-0.5$ a.u. 
However, in the spin-compensated stretched H$_2$ both the spin-up and the 
spin-down (noninteracting) KS electrons are found with a 50\% chance on either 
nucleus. Thus what enters as the atomic term $\chi^{0}[\mbox{H}]$ in Eqs.~\ref{eq:chiksh2} and \ref{eq:chilambdaat} corresponds in fact
to the KS spin-response of a 
fictitious spin-compensated H atom (denoted H') with half occupied $1s\!\uparrow$ 
and $1s\!\downarrow$ states. The corresponding RPA+X correlation energy appearing on the R.H.S. of Eq.~\ref{eq:ech2infty} is therefore
$E\c^{\textrm{RPA+X}}[\mbox{H'}] \simeq E\c^{\textrm{RPA}}[\mbox{H}]/2$ (estimated to second order in $\lambda v_{\textrm{ee}}$). Consequently, $\lim_{R\rightarrow\infty}
(E_{\rm tot}^{\textrm{RPA+X}}[\mbox{H}_2] - E_{\rm tot}^{\textrm{RPA+X}}[\mbox{H}]) 
\simeq E\c^{\textrm{RPA}}[\mbox{H}] < 0$. In an actual calculation 
we indeed find that the RPA+X potential energy curve drops below zero 
beyond $R\simeq 7$~bohr (see Fig.~\ref{fig:pes}).

The RPA yields $E\c^{\textrm{RPA}}[\mbox{H'}]=E\c^{\textrm{RPA}}[\mbox{H}]$,
i.e. the distinction between H' and H does not matter. This (spurious) RPA self-correlation 
energy ($E_c^{\textrm{RPA}}[\mbox{H}] \approx -23$~mHa per atom, using the exact $n_{\textrm{H}}$) appears for the free
atoms and the stretched H$_2$ and thus cancels out. In a spin-polarized one-electron system, 
the exchange kernel fully cancels the Coulomb kernel, leading to $E\c^{\textrm{RPA+X}}[\mbox{H}]=0$. 
In a spin-compensated two-electron system like H$_2$, correlation should arise only from
the interaction of opposite-spin electrons.
For the RPA+X kernel it is easy to see that there is zero like-spin correlation
up to second order in $\lambda v_{\textrm{ee}}$ (although non-zero in higher orders). The fact that the
dynamical correlation associated with $\tilde\chi^{\lambda}$ does not
vanish for the RPA+X kernel in the infinitely stretched H$_2$ may then be seen as a failure
to suppress opposite-spin correlation between the electrons on either atomic site.

Had we treated the stretched H$_2$ in a spin-polarized KS scheme, the two electrons
would be localized with opposite spin on either nucleus right from the beginning.
While we have not performed this calculation, we expect from our above discussion
that both the RPA and RPA+X yield  $\lim_{R\rightarrow\infty}E_{\textrm{tot}}[\mbox{H}_2]=2E_{\textrm{tot}}[\mbox{H}]$ in this case.

{\em Model based on the HOMO-LUMO transition only:} 
If we include in the response only the HOMO and LUMO 
KS states, we get a ``minimal'', two-state model of H$_2$
in which all effects of higher transitions are ignored.
Then $E\c[\mbox{H}_2]$ in Eq.~\ref{eq:ech2infty} is given by 
just the static correlation term $\delta E\c[\mbox{H}_2]$: 
\begin{equation}
\begin{array}{rcl}
E\c[\mbox{H}_2] &\sim& \frac{E_g}{\kappa}
	\left(
        \sqrt{1+\frac{2\kappa K_0}{E_g}} - 1
	\right) - K_0 
\\
&\simeq&
                -K_0 + \sqrt{2K_0 E_g/\kappa}\quad,\quad\mbox{for}\,R\rightarrow\infty\quad,
\label{eq:decmin}
\end{array}
\end{equation}
for the RPA ($\kappa=2$) as well as the RPA+X ($\kappa=1$). The
asymptotic $R$-dependence of this minimal $E\c[\mbox{H}_2]$ is again that given by Eq.~\ref{eq:decrpainf}.
Hence within the minimal model also the RPA+X, like the RPA, correctly
yields the total energy of the dissociated H$_2$ as that of two free 
H atoms.
Indeed, inspection of Eq.~\ref{eq:deluc} shows that both RPA and
RPA+X recover the exact adiabatic connection for $R\rightarrow\infty$,
i.e. that $\delta U_{\textrm{c}}(\lambda > 0)=-U[\mbox{H}]$.
Note however that due to the closing
HOMO-LUMO gap for $R\rightarrow\infty$ the initial slope
$dU\c^{\textrm{RPA+X}}(\lambda)/d\lambda|_{\lambda=0}
=-K_0^2/E_g$ eventually diverges, as does the the GL2 correlation energy.
Independently of our work, an analogous minimal model has been recently
obtained by van Leeuwen and coworkers, studying total energy 
functionals based on Green functions~\cite{com:dlb}.


\begin{thebibliography}{99}
\bibitem{HK64}{
        P. Hohenberg and  W. Kohn, 
	Phys. Rev. {\bf 136}, B864 (1964).
}
\bibitem{KS65}{
	W. Kohn and  L.J. Sham,
        Phys. Rev. {\bf 140}, A1133 (1965).
}
\bibitem{kh01a}{
	W. Koch and M.C. Holthausen, {\it A Chemist's Guide to
	Density Functional Theory} (Wiley-VCH, Weinheim, 2001).
}
\bibitem{bec93a}{
        A.D. Becke,
        J. Chem. Phys. {\bf 98}, 5648 (1993);
        {\it ibid.} {\bf 104}, 1040 (1996).
}
\bibitem{per96a}{
        J.P. Perdew, M. Ernzerhof, and K. Burke,
        J. Chem. Phys. {\bf 105}, 9982 (1996).
}
\bibitem{dob94a}{
	J. Dobson, in ``Topics of Condensed Matter Physics'',
	edited by M.P. Das (Nova, New York 1994), p. 121;
	see also \url{cond-mat/0311371}.
}
\bibitem{dob96a}{
        J.F. Dobson and B.P. Dinte,
        Phys. Rev. Lett. {\bf 76}, 1780 (1996).
} 
\bibitem{koh98a}{
	W. Kohn, Y. Meir, and D.E. Makarov,  
	Phys. Rev. Lett. {\bf 80}, 4153 (1998).
}
\bibitem{eng00b}{
	E. Engel, A. H\"ock, and R.M. Dreizler,
	Phys. Rev. A {\bf 61}, 032502 (2000).
}
\bibitem{ggg02}{
	P. Garc\'ia-Gonz\'ales and R.W. Godby,
	Phys. Rev. Lett. {\bf 88}, 056406 (2002).
}
\bibitem{ern97a}{
	M. Ernzerhof, K. Burke, and J.P. Perdew, in:
	{\it Recent Developments in Density Functional Theory},
	edited by J.M. Seminario (Elsevier, Amsterdam, 1997).
}
\bibitem{bal97a}{
	T. Bally and G. Nahsari Sastry,
	J. Phys. Chem. A {\bf 101}, 7923 (1997).
}
\bibitem{sod99a}{
	M. Sodupe, J. Bertran, L. Rodr{\'i}guez-Santiago, and
	E.J. Baerends, 
	J. Phys. Chem. A {\bf 103}, 166 (1999).
}
\bibitem{ern97b}{
	M. Ernzerhof, J.P. Perdew, and K. Burke, in:
        {\it Topics in Current Chemistry}, Vol. 180,
	edited by R.F. Nalewajski (Springer, Berlin, 1996), p. 1.
}
\bibitem{gri97a}{
	O.V. Gritsenko, P.R.T. Schipper, and E.J. Baerends,
	J. Chem. Phys. {\bf 107}, 5007 (1997).
}
\bibitem{per95a}{
	J.P. Perdew, A. Savin, and K. Burke,
	Phys. Rev. A {\bf 51}, 4531 (1995).
}
\bibitem{lee93a}{
	A.M. Lee and N.C. Handy,
	J. Chem. Soc. Faraday Trans. {\bf 89}, 3999 (1993).
}
\bibitem{gor99a}{
	C.D. Sherill, M.S. Lee, and M. Head-Gordon,
	Chem. Phys. Lett. {\bf 302}, 425 (1999).
}
\bibitem{fil00a}{
	M. Filatov and S. Shaik,
	Chem. Phys. Lett. {\bf 332}, 409 (2000);
	J. Phys. Chem. A {\bf 104}, 6628 (2000).
}
\bibitem{pol02a}{
	R. Pollet, A. Savin, T. Leininger, and H. Stoll,
	J. Chem. Phys. {\bf 116}, 1250 (2002).
}
\bibitem{lan75a}{
        D.C. Langreth and J.P. Perdew,
        Solid State Commun. {\bf 17}, 1425 (1975);
        Phys. Rev. B {\bf 15}, 2884 (1977).
}
\bibitem{perdew01p}{
	J.P. Perdew and K. Schmidt,
	in: {\it Density Functional Theory and Its Application to Materials},
	edited by V. Van Doren, C. Van Alsenoy, and P. Geerlings,
	(AIP, Melville, New York, 2001).
}
\bibitem{kur99a}{
        S. Kurth and J.P. Perdew,
        Phys. Rev. B {\bf 59}, 10461 (1999).
}
\bibitem{dob00a}{
        J.F. Dobson and J. Wang,
        Phys. Rev. Lett. {\bf 82}, 2123 (1999).
}
\bibitem{fur01a}{
	F. Furche,
	Phys. Rev. B {\bf 64}, 195120 (2001).
}
\bibitem{fuc02a}{
	M. Fuchs and X. Gonze,
	Phys. Rev. B {\bf 65}, 235109 (2002).
}
\bibitem{ryd03a}{
	H. Rydberg {\it et al.}, 
	Phys. Rev. Lett. {\bf 91}, 126402 (2003).
}
\bibitem{dio04a}{
	M. Dion, H. Rydberg, E. Schr\"oder, D.C. Langreth, and B.I. Lundqvist,
	Phys. Rev. Lett. {\bf 92}, 246401 (2004).
}
\bibitem{yan00a}{
	Z. Yan, J.P. Perdew, and S. Kurth,
	Phys. Rev. B {\bf 61}, 16430 (2000).
}
\bibitem{dob00b}{
	J.F. Dobson and J. Wang,
	Phys. Rev. B {\bf 62}, 10038 (2000).
}
\bibitem{dob02a}{
	J.F. Dobson, J. Wang, and T. Gould,
	Phys. Rev. B {\bf 66}, 081108 (2002).
}
\bibitem{gru03a}{
	M. Gr{\"u}ning, O.V. Gritsenko, and E.J. Baerends,
	J. Chem. Phys. {\bf 118}, 7183 (2003).
}
\bibitem{her03a}{
	J.M. Herbert and J.E. Harriman,
    	Chem. Phys. Lett. {\bf 382}, 142 (2003).
}
\bibitem{bae01a}{
	E.J. Baerends, 
	Phys. Rev. Lett. {\bf 87}, 133004 (2001).
}
\bibitem{kol60a}{
        W. Kolos and C.C. Roothaan,
        Rev. Mod. Phys. {\bf 32}, 219 (1960).
}
\bibitem{com:ymn}{
        Of course, one never works with this decomposition within DFT. The 
        interpretation in terms of covalent and ionic parts is however
        quite natural when $\Psi^{\rm KS}$ is analyzed as a many-electron
        wavefunction.
}
\bibitem{cou}{C.A. Coulson and I. Fisher, Phil. Mag. {\bf 40}, 386 (1949).
}
\bibitem{gun76a}{
	O. Gunnarsson and B. Lundqvist,
	Phys. Rev. B {\bf 13}, 4274 (1976).
}
\bibitem{leeuwen96}{
	R. van Leeuwen, O.V. Gritsenko, and E.J. Baerends
	in: {\it Topics in Current Chemistry}, Vol. 180,
        edited by R.F. Nalewajski (Springer, Berlin, 1996), p. 107.
}
\bibitem{fil02a}{
	C. Filippi, S.B. Healy, P. Kratzer, E. Pehlke, 
	and M. Scheffler, 
	Phys. Rev. Lett. {\bf 89}, 166102 (2002). 
}
\bibitem{bau96a}{
	R. Bauernschmitt and R. Ahlrichs,
        J. Chem. Phys. {\bf 104}, 9047 (1996). 
}
\bibitem{gra98a}{
	J. Gr\"afenstein, E. Kraka, and D. Cremer,
	Chem. Phys. Lett. {\bf 288}, 593 (1998).
}
\bibitem{gri99a}{
	S. Grimme and M. Waletzke,
	J. Chem. Phys. {\bf 111}, 5645 (1999).
}
\bibitem{bae99f}{
	P.R.T. Schipper, O.V. Gritsenko, and E.J. Baerends,
	J. Chem. Phys. {\bf 111}, 4056 (1999).
}
\bibitem{goe94a}{
        A. G\"orling and  M. Levy,
        Phys. Rev. A {\bf 50}, 196 (1994); 
	for a review see also:
	T. Grabo, T. Kreibich, S. Kurth, and E.K.U. Gross,
	in: {\it Strong Coulomb Correlations in Electronic
	Structure: Beyond the LDA}, edited by V.I. Anisimov
	(Gordon and Breach, 1999), p. 203.
}
\bibitem{pet96a}{
        M. Petersilka, U.J. Gossmann, and E.K.U. Gross,
        Phys. Rev. Lett. {\bf 76}, 1212 (1996).
}
\bibitem{goe93a}{
	A. G\"orling and M. Levy,
	Phys. Rev. B {\bf 47}, 13105 (1993).
}
\bibitem{bec98a}{
	A.D. Becke, 
        J. Chem. Phys. {\bf 109}, 2092 (1998).
}
\bibitem{scu98a}{
	T. Van Voorhis and G.E. Scuseria, 
	J. Chem. Phys. {\bf 109}, 400 (1998).
}
\bibitem{tao03a}{
	J. Tao, J.P. Perdew, V.N. Staroverov, and G.E. Scuseria
	Phys. Rev. Lett. {\bf 91}, 146401 (2003). 
}
\bibitem{alm99a}{
	C.-O. Almbladh, U. von Barth, and R. van Leeuwen,
	Int. J. Mod. Phys. B {\bf 13}, 535 (1999).
}
\bibitem{dah04a}{
	N.E. Dahlen and U. von Barth, 
	J. Chem. Phys. {\bf 120}, 6826 (2004).
}
\bibitem{hol90}{
	L.J. Holleboom and J.G. Snijders,
	J. Chem. Phys. {\bf 93}, 5826 (1990).
}
\bibitem{ary01a}{
	F. Aryasetiawan, T. Miyake, and K. Terakura,
	Phys. Rev. Lett. {\bf 88}, 166401 (2002);
	{\it ibid.} {\bf 90}, 189702 (2003).
}
\bibitem{fuc03a}{
	M. Fuchs, K. Burke, Y.-M. Niquet, and X. Gonze,
        Phys. Rev. Lett. {\bf 90}, 189701 (2003).
}
\bibitem{niq03a}{
	Y.-M. Niquet, M. Fuchs, and X. Gonze,
	J. Chem. Phys. {\bf 118}, 9504 (2003).
}
\bibitem{niq03b}{
	Y.-M. Niquet, M. Fuchs, and X. Gonze,
	Phys. Rev. A {\bf 68}, 032507 (2003).
}
\bibitem{xclda}{
	J.P. Perdew and Y. Wang,
        Phys. Rev. B {\bf 45}, 13244 (1992).
}
\bibitem{xcpbe}{
	J.P. Perdew, K. Burke, and  M. Ernzerhof,
        Phys. Rev. Lett. {\bf 77}, 3685 (1996).
}
\bibitem{lev85c}{
	M. Levy and J.P. Perdew, 
	Phys. Rev. A {\bf 32}, 2010 (1985).
}
\bibitem{ern96a}{
        M. Ernzerhof, 
        Chem. Phys. Lett. {\bf 263}, 499 (1996).
}
\bibitem{BEP97}{
	K. Burke, M. Ernzerhof, and J.P. Perdew,
	Chem. Phys. Lett. {\bf 265}, 115 (1997).
}
\bibitem{rvlphd}{
	R. van Leeuwen,
	Ph.D. thesis, Vrije Universiteit, Amsterdam (1996).
}
\bibitem{gri96a}{
	O.V. Gritensko and E.J. Baerends,
	Phys. Rev. A {\bf 54}, 1957 (1996).
}
\bibitem{com:interpolate}{
	We interpolate the (unknown) exact curve in $\lambda\in[0,1]$ 
        by $\Delta U\xc(\lambda)=\Delta E\x + 
	\frac{\Delta U\xc'(0)\lambda p(\lambda)}{1+\alpha\lambda p(\lambda)}$ 
        where $\alpha=\frac{\Delta U\xc'(0)}{\Delta U\c}-\frac{1}{p(1)}$. 
	Our interpolation reproduces $\Delta E\x$, $\Delta U\xc'(0)$,
	and $\Delta U\xc$ as given in Tabs.~\ref{tab:aceq} and \ref{tab:ac5}. 
	For $R=1.4$~bohr it integrates to $\Delta E\xc=-2.07$~eV, where we 
	have set $p(\lambda)\equiv 1$.   
	For $R=5$~bohr we have chosen $p(\lambda)=1+\beta\lambda e^{-\gamma\lambda}$, 
	adjusting $\beta$ and $\gamma$ to yield $\Delta E\xc=0.82$~eV and also
	the strong interaction limit of $U\xc[\mbox{H}_2](\lambda)$ [M. Seidl, private communication].
} 
\bibitem{puz01a}{
	A. Puzder, M.Y. Chou, and R.Q. Hood,
	Phys. Rev. A {\bf 64}, 022501 (2001).
}
\bibitem{sav01a}{
	A. Savin, F. Colonna, and M. Allavena,
        J. Chem. Phys. {\bf 115}, 6827 (2001).
}
\bibitem{sav03a}{
	F. Colonna, D. Maynau, and A. Savin,
	Phys. Rev. A {\bf 68}, 012505 (2003).
}
\bibitem{hoo98a}{
	R.Q. Hood, M.Y. Chou, A.J. Williamson, G. Rajagopal, and R.J. Needs,
	Phys. Rev. B {\bf 57}, 8972 (1998).
}
\bibitem{mtb03}{
	R.J. Magyar, W. Terilla, and K. Burke, 
	J. Chem. Phys. {\bf 119}, 696 (2003).
} 
\bibitem{sei00a}{
	M. Seidl, J.P. Perdew, and S. Kurth,
	Phys. Rev. Lett. {\bf 84}, 5070 (2000).
}
\bibitem{com:isi}{
	M. Fuchs and X. Gonze, unpublished.
}
\bibitem{ada99a}{
	C. Adamo and V. Barone,
	J. Chem. Phys. {\bf 110}, 6158 (1999).
}
\bibitem{cle90a}{
	E. Clementi and S.J. Chakravorty, 
	J. Chem. Phys. {\bf 93}, 2591 (1990).
}
\bibitem{umr94a}{
	C.J. Umrigar and  X. Gonze,
	Phys. Rev. A {\bf 50}, 3827 (1994).
}
\bibitem{fuc02b}{
	M. Fuchs and X. Gonze, unpublished.
}
\bibitem{lei99a}{
	M. Lein, E.K.U. Gross, and J.P. Perdew,
	Phys. Rev. B {\bf 61}, 13431 (2000).
}
\bibitem{cas96a}{
	M.E. Casida, 
	in {\it Recent Developments and Applications in 
	Density Functional Theory}, edited by J.M. Seminario
	(Elsevier, Amsterdam 1996).
}
\bibitem{slater}{
	J.C. Slater, 
	Quantum Theory of Molecules and Solids, Vol. 1
	(McGraw-Hill, New York, 1963), p. 60.
}
\bibitem{gri00a}{
	O.V. Gritsenko, S.J.A. van Gisbergen, A. G\"orling, and E.J. Baerends,
	J. Chem. Phys. {\bf 113}, 8478 (2000).
}
\bibitem{mzcb04}{
	N.T. Maitra, F. Zhang, R.J. Cave, and K. Burke,
	J. Chem. Phys. {\bf 120}, 5932 (2004).
}
\bibitem{abinit}{
	X. Gonze {\it et al.}, 
	Comput. Materials Science {\bf 25}, 478 (2002).
}
\bibitem{fuc99a}{
	M. Fuchs and M. Scheffler,
	Comput. Phys. Commun. {\bf 107}, 67 (1999).
}
\bibitem{AGB03}{
        H. Appel, E.K.U. Gross, and K. Burke, 
	Phys. Rev. Lett. {\bf 90}, 043005 (2003).
}
\bibitem{com:dlb}{
        R. van Leeuwen, private communication.
}
\end{thebibliography}
\end{document}